\documentclass[a4paper,11pt]{article}

\usepackage[utf8]{inputenc}
\usepackage{cancel}
\usepackage{tikz}
\usepackage{ulem}
\usepackage{amsfonts}
\usepackage{amssymb}
\usepackage{graphicx}
\usepackage{amsmath}
\usepackage{enumerate}
\usepackage{mathtools}
\usepackage{subfig}
\usepackage{color}
\usepackage{tikz}
\usepackage{float}
\usepackage{here}
\usepackage{cite}
\usepackage{mathrsfs}
\usepackage{float,epsfig}
\usepackage{dcolumn}
\usepackage{graphicx}
\usepackage{bm}
\usepackage{booktabs}
\usepackage{multirow}
\usepackage{changepage}
\usepackage{amsmath,amssymb,amsthm}
\usepackage[colorlinks=true,linkcolor=blue,citecolor=red]{hyperref}
\usepackage{booktabs} 
\usepackage{siunitx}  
\tikzset{every picture/.style={line width=0.75pt}} 
\usetikzlibrary{arrows.meta, positioning}
\usepackage{multirow}
\usepackage[toc,page]{appendix}
\usetikzlibrary{arrows.meta}
\usetikzlibrary{bending}
\usetikzlibrary{calc}
\newcommand{\be}{\begin{equation}}
\newcommand{\ee}{\end{equation}}
\newcommand{\bea}{\setlength\arraycolsep{2pt} \begin{eqnarray}}
\newcommand{\eea}{\end{eqnarray}}

\def\0{{\sst{(0)}}}
\def\1{{\sst{(1)}}}
\def\2{{\sst{(2)}}}
\def\3{{\sst{(3)}}}
\def\4{{\sst{(4)}}}
\def\5{{\sst{(5)}}}
\def\6{{\sst{(6)}}}
\def\7{{\sst{(7)}}}
\def\8{{\sst{(8)}}}
\def\sst#1{{\scriptscriptstyle #1}}

\makeatletter \@addtoreset{equation}{section}

\definecolor{lime}{HTML}{A6CE39}

\begin{document}
\title{{\normalsize \textbf{\Large  Learning  
 Parameters  of   dRGT Black
Holes in   Nonlinear Electrodynamics with    Neural Network   Scenarios}}}
\author{ {\small   Adil Belhaj$^1$, Salah Eddine Ennadifi$^2$,   Maryem  Jemri$^1$
 \footnote{
Authors are listed  in alphabetical order. 
} \hspace*{-8pt}} \\
{\small  $^1$ESMaR, Faculty of Science, Mohammed V University in Rabat, Rabat, Morocco  }\\
{\small  $^2$LHEP-MS,   Faculty of Science, Mohammed V University in Rabat, Rabat, Morocco }
}
\maketitle

\begin{abstract}
Using machine learning  methods,  we  present   constraining   parameter scenarios     of  AdS black holes in dRGT-like massive gravity coupled to nonlinear electrodynamics.  We first  approach   the photon sphere, the  black hole shadow, and the  energy emission rate  with the help of  high  performance parallel computations.     To corroborate   such optical    results, we  develop a numerical  framework to   constrain the  electric charge  and the  nonlinear electrodynamic  parameter,  for certain   gravity regimes,  via    the shadow   observations reported by the Even Horizon Telescope collaboration.  Employing neural networks, we then examine the parameter regions of  the black holes based on their consistency with EHT observations of M87$^*$, SgrA$^*_{\mathrm{VLTI}}$, and SgrA$^*_{\mathrm{Keck}}$, within the $1-\sigma$ and $2-\sigma$ confidence intervals.   As a result,  we find that the obtained constraints  are consistent with  the $SgrA^*_{\mathrm{VLTI}}$  empirical  data.

\end{abstract}

\newpage

\section{Introduction}
Black hole thermodynamics has been extensively   studied     by implementing    many  gravity theories.  Various   quantities, such as entropy and temperature, have been  computed and examined   via  different  approaches  and methods  using analytical and numerical  computations.     A particular emphasis has been put   on     the  investigation of  the  phase transition and  the critical behaviors \cite{ad1,ad2, 01, ad3, ad30, ad31,18}.   It has been observed that  such activities  have unveiled  relevant data needed to understand the physics  around   black hole  objects. This has established   primordial   bridges  between different  theories,  topics,   and   branches of physics. Precisely,  it has been provided links between  Type II  superstring theory, D-brane physics, quantum information theory,  advanced mathematics dealing with non-trivial  geometries like Calabi-Yau manifolds, and computer sciences with non-trivial  numerical computations including  parallel  programing simulations. 

According to black hole physics activities, the 
optical  behaviors  have received  a special  interest supported by recent empirical findings  \cite{601,70,80, 81,90,100,0001}. In fact,  the light    in 
strong gravitational  dynamics  can   supply relevant 
data concerning the spacetime geometry  with the help of 
 certain  analytical and numerical  approaches.   These studied   have   shown certain observable effects such as   the 
gravitational lensing and  the  shadow configurations in terms of  real curves in  the plane.  Moreover, the Event Horizon Telescope (EHT)   international collaborations   have  provided  relevant developments   to test many theories which could generate 
observational black hole models \cite{6,7}.   These  activities  have been supported  by  
the first images of the shadows of supermassive black 
holes reported by EHT.  These  observational results    have paved new gates   to check the obtained  theoretical model of  the  ordinary  gravity theory. Based on such developments,   several links  between  the black hole physics and  the astrophysical  observational results have been established advancing   novel methods.

Such  black hole physics  developments have  motivated    the  investigation  of  certain gravity theories extending  the ordinary  solutions.    A close examination shows  that  modified theories of gravity  coupled to generalized  electromagnetic  field models have been largely studied.  These  extended  models  have been explored   to approach  issues  appearing in  cosmology and high-energy physics  \cite{mod1}. Many models have been dealt with including  de Rham--Gabadadze--Tolley (dRGT) massive gravity  supplying a relevant framework \cite{555,565}. In  this   theory, the graviton   can gain  a non-zero mass  $m_g$  producing  non-trivial gravitational  effects.   Concretely, dRGT-like  models involving  singular  metrics have been studied  generating non-trivial   black hole solutions  controlled by   external  massive gravity parameters  producing  extra regions in the involved  moduli space  \cite{57}.   At has been observed that such  parameters can significantly affect   the spacetime geometry   modifying certain  physical behaviors  including  the  thermodynamic universalities    and the  phase structures \cite{58}. 

Recently, the thermodynamics of charged AdS black holes in dRGT-like massive gravity with  nonlinear electrodynamics  has  been  studied.   Concretely,   the ModMax electrodynamic contributions  have  been  utilized   to approach  certain issues associated with  the Joule--Thomson expansion,  the $P$--$V$ criticality, and  the Gibbs free energy. The  corresponding findings   have provided   thermodynamic aspects.  Precisely,   it has been observed  that the Van der Waals-like phase transitions and nontrivial cooling regimes have been  investigated  \cite{88,880}. Alternative  studies   have been reported  to examine aspects  via      exponential nonlinear electrodynamics in dRGT massive gravity \cite{99}.  It has been shown that  the graviton mass and the electromagnetic nonlinearities  have impacted   the thermodynamic phase structure behaviors.  These explored activities  can be exploited to address    questions dealing with   black hole properties including  the  optical ones  and  the observational signature features. 

More recently, machine learning  techniques   have applied  to   black hole physics.  This  method  has generated efficient tools  to   analysis,   complex,  rational and empirical  data of  black hole  physics.  Precisely,  machine learning  approaches   have been  exploited  to estimate the relevant  physical parameters. They have  also been  used to  classify different black hole configurations.  Moreover, such numerical techniques   have improved the reconstruction and  the interpretation of black hole  pictures.    Indeed, it has  been shown that neural networks (NNs)   have been  used  to deduce the   accretion parameters from  the horizon-scale images \cite{14}. Moreover,  the black hole images obtained by the EHT  international collaborations have been improved significantly  \cite{15}.  Alternatively,  machine learning methods  have been used to classify black holes in the mass--spin parameter space. Examinations  have  revealed  the potential of such numerical  methods to  derive certain physical data which can be exploited to   investigate the   black hole behaviors \cite{16}. Precisely, many aspects of these objects have been approached via machines  learning methods providing relevant results \cite{160,161,162,163,164,165}.

The aim of this work is  to  study constraining parameter scenarios    of  AdS  black  holes in  dRGT-like massive gravity with   nonlinear  electrodynamics using machine learning techniques via shadow computations.   Precisely, we  first   deal with  the   optical properties of non-rotating  solutions with  ModMax nonlinear electrodynamic contributions by approaching  the photon sphere, the  black hole shadow, and the energy emission rate for different values of the massive gravity parameters  using high-performance parallel computations.  To  support the  obtained  results with the shadow observations reported by the EHT  international collaborations, we then  develop a numerical code  through  parallel programing  computations to   constrain the  electric charge $Q$ and the ModMax parameter $\gamma$.   Exploiting  such  numerical data, we  build  a training dataset for an NN  used to  determine   the  black hole parameter regions being  compatible with the EHT observations of M87$^*$, SgrA$^*_{\mathrm{VLTI}}$, and SgrA$^*_{\mathrm{Keck}}$ within the $1-\sigma$ and $2-\sigma$ confidence intervals.  As a result,  we find that the obtained constraints  match with the SgrA$^*_{\mathrm{VLTI}}$  empirical  findings.

This  paper is organized  as follows. In section 2, we give  a concise discussion on   the behaviors  of non-rotating AdS black holes in dRGT-like massive gravity coupled to the  ModMax nonlinear electrodynamics. In  section 3,  we study  the photon sphere,  the black hole shadow, and  the energy emission rate using advanced numerical techniques  based on parallel programing simulations. Then,  we  confront the theoretical predictions with observational data of  the EHT international  collaborations. In  Section 4,  we use   machine learning techniques   to  determine whether the resulting parameter space are   consistent with the observations of $M87^*$, $SgrA^*_{\mathrm{VLTI}}$, and $SgrA^*_{\mathrm{Keck}}$ models.   In the last section, we  give certain concluding remarks.
\section{AdS black holes in   dRGT-like massive gravity with  non-linear electrodynamics}
Massive gravity extends general relativity by giving the graviton a nonzero
mass $m_{g}>0$.  A massive graviton propagates extra polarization states and
mediates a Yukawa-type interaction of range $m_{g}^{-1}$. \ The
corresponding consistent action of this idea, provided by dRGT theory \cite%
{555,565}, reads, roughly, as

\begin{equation}
S=\frac{1}{\sqrt{16\pi G}}\int d^{4}x\sqrt{-g}\left[ R+m_{g}^{2}U+\zeta
_{matter}\right]   \label{ee}
\end{equation}%
where $R$ is the Ricci scalar, $m_{g}^{2}U$ is the graviton mass term where  $U
$ is the graviton potential being a metric function indicating how the
physical metric $g_{\mu \nu }$ differs from the reference metric $f_{\mu \nu
}$. In  four dimensions,  for instance, such a potential has the particular form 
$U=U\left( g_{\mu \nu },f_{\mu \nu }\right) =\sum\limits_{n=0}^{4}c_{n}U_{n}
$ where  the dimensionless coefficients $c_{n}$ are arbitrary constants
corresponding to different massive gravity models. Coupling this sector to the 
ModMax nonlinear electrodynamics yields black hole solutions controlled by
these massive-gravity parameters $c_{n}$.  Roughly, we  present essential materials concerning the   AdS black holes in dRGT-like massive gravity  with   nonlinear electrodynamics \cite{60}.    Indeed,   we  take  a static and  spherically symmetric line element with the following form
\begin{equation}
ds^{2}
=
-f(r)\,dt^{2}
+\frac{dr^{2}}{f(r)}
+r^{2}d\theta^{2}
+r^{2}\sin^{2}\theta\,d\phi^{2},\label{ff}
\end{equation}
where $f(r)$  is the metric   function carrying  data on the effect of  the extra background contributions on the ordinary AdS black holes. This radial  function depends on the   external parameters  modifying the thermodynamic and  the optical properties of  the considered  black holes. These modifications usually  extend  the   moduli space which can be exploited to bring appropriate universal rations   by considering specific  parameters and exploiting EHT empirical  results. Supported by the results of \cite{60} and forgetting about  the action calculations  deriving the  solutions reported in  the associated references, we   consider  the following metric function form 
  \begin{equation}
  f(r)=1-\frac{2M}{r}-\frac{\Lambda}{3}r^{2}
+\frac{e^{-\gamma}Q^{2}}{r^{2}}
+m_{g}^{2}C\left(\frac{c_{1}r}{2}+c_{2}C\right)
  \end{equation}
where  $M$ and $Q$  are  the mass and the charge  of the black hole, respectively.
The dimensionless parameter $\gamma$  which is positive or zero  describes the ModMax nonlinear electrodynamics. The terms proportional to $m_g^2$  are  originated  from the massive graviton sector.    $c_{1}$ and $c_{2}$ are  the only  considered  constants characterizing  the massive graviton potential. However,  $C$ is a positive constant determined by the spatial reference metric.  Taking  $\gamma=0 $, one recovers  the standard Maxwell electrodynamics.   Sending  $m_{g}$ to zero, however, one obtains  the ordinary Reissner--Nordström--AdS (RN--AdS) black hole. To explore the optical properties of the  proposed black hole solutions, we will  perform  accelerated  parallel numerical simulations to   approach   the photon sphere,  the black hole shadows, and the  energy emission rate. Then, we  establish   constraint scenarios reducing corroborated regions of the moduli space.   For a  comparative study, we consider two representative dRGT-like massive gravity models being  a weak-coupling case $(m_g=0.4, C=0.1, c_1=1, c_2=0.2) $ and a strong coupling case $(m_g=0.8, C=0.4, c_1=-5, c_2=7)$.  

Having  approached  the metric function, we  discuss now   the horizon structure with the help of advanced numerical  computations,  enabling efficient parallel computations on GPUs \cite{v,46}.
 The event horizons  can be identified by determining  the metric function roots. In fact, this can be obtained  by solving the  algebraic equation corresponding to the vanishing of the radial metric component.

To  determine the existence of the possible  horizons, we  develop   numerical algorithms for both the weak and  the strong massive gravity regimes. The  black hole parameters $Q$ and $\gamma$  vary independently over the interval $[0,1]$ with a step size of $0.01$. Considering  each couple $(\gamma,Q)$,  the horizon equation can be  approached  numerically  in order to  precise whether at least one real positive root exists,  representing  a physical  solution.

\begin{figure}[ht!]
\centering
\begin{tabular}{cc}
\includegraphics[width=0.3\linewidth]{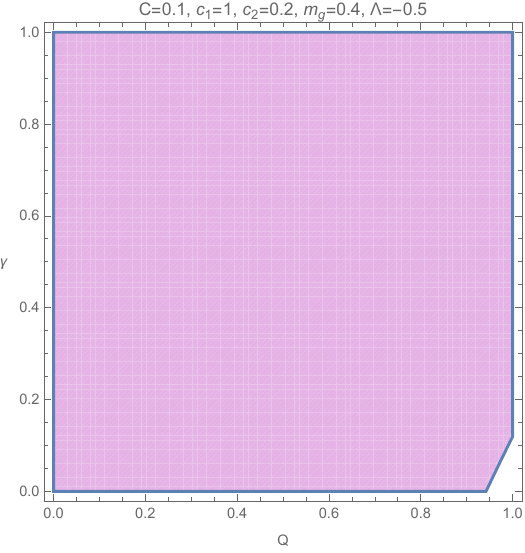} &
\includegraphics[width=0.3\linewidth]{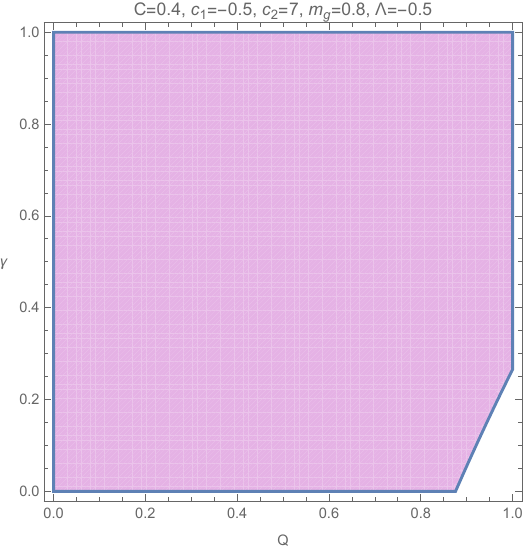}
\end{tabular}
\caption{ Regions in the $(\gamma,Q)–$plane, where the metric admits at least one real event horizon
radius.}
\label{fig:11}
\end{figure}

Before going ahead, we would like to  identify the regions of  the moduli  space where the black hole solution admits  the physical horizons. Roughly,  Fig.~\ref{fig:11} represents the resulting parameter space distributions  showing   the regions in the $(\gamma,Q)$ plane where at least one event horizon exists for both the weak and  the strong massive gravity regimes.
 For small values of $\gamma$ and large values of $Q$, Fig.~\ref{fig:11} reveals  that the physical horizon solutions belong to  a very small region of the  reduced moduli space $(\gamma, Q)$.  Moreover, the weak massive gravity regime generates regions of  the existence being  relevant to the physical horizons relative to the strong massive gravity regime. 
 
 In what follows,   we investigate the corresponding  optical behaviors by    considering  regions belonging to  the moduli parameter space matching with the allowed domain admitting at least one real solution.
\section{ Shadow curve configurations   and EHT observations}

In this section, we  study the optical  behaviors  of charged AdS black holes in dRGT-like massive gravity coupled to the  nonlinear electrodynamics.  Precisely,  we approach the photon sphere, the black hole shadow, and the energy emission rate.  To do so, one should  study  null geodesics playing a fundamental role in the  understanding of  the  light propagation  in the strong gravitational fields   around  black hole objects.  This procedure  unveil  valuable  data concerning   relevant observable phenomena,  such as  the  black hole shadows,  the gravitational lensing, and  the photon trajectories.  Such  optical signatures constitute powerful tools not only  to  probe  the underlying spacetime geometry but also to  test both general relativity and its modified extensions \cite{73,74}.  In this way,  the first horizon-scale images of the supermassive black holes M87$^{*}$ and Sgr A$^{*}$ obtained by the EHT have motivated extensive  activities   dealing with such   photon motion scenarios  in    black hole geometries. Within this framework, we can  derive the conditions governing the existence of the photon sphere and determine the associated  radius for the studied   black hole solutions.   Moreover,  we  study  the energy emission rate, being  closely related to the absorption cross section in the geometric optics limit  supplying   further insights into the thermodynamics and  the  observational  optical   aspect.
\subsection{ Shadow computations }
\begin{figure}[ht!]
\centering
\begin{tabular}{cc}
\includegraphics[width=0.5\linewidth]{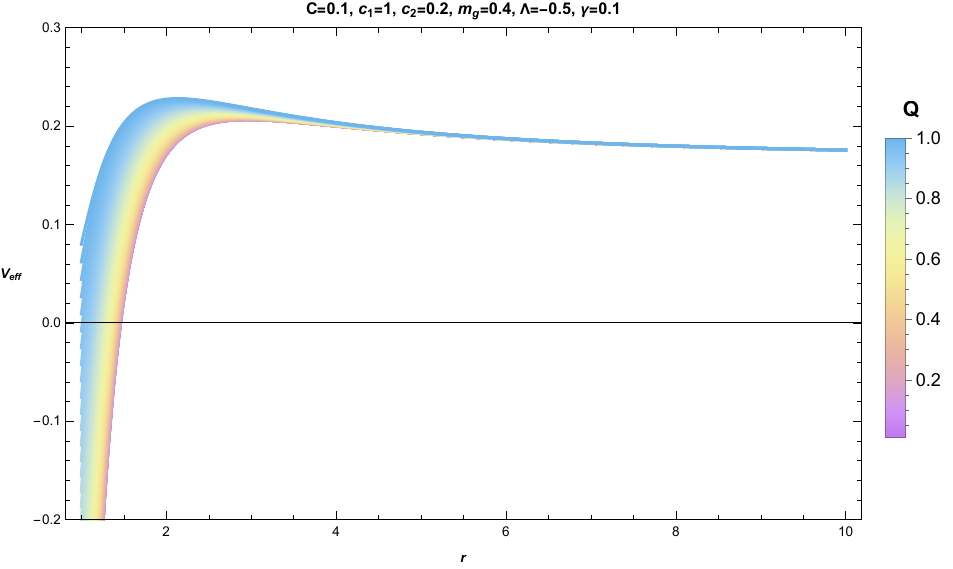} &
\includegraphics[width=0.5\linewidth]{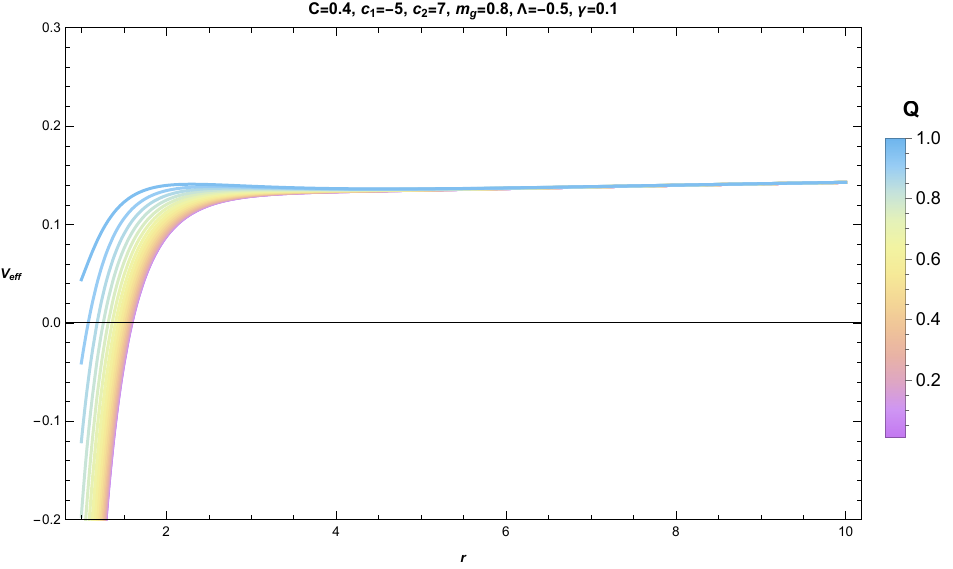}
\end{tabular}
\caption{Behavior of the effective potential $V_{\rm eff}$ as a function of the radial coordinate $r$ for  weak and strong massive gravity.}
\label{fig:1}
\end{figure}

To start this  investigation,  we first   need  to  establish the geodesic equations via the Lagrangian formalism,  providing  a systematic framework to  study  the trajectories of photons in curved spacetime geometries.  In this way, the  Lagrangian of a test particle in the presence of  a gravitational field can be determined using the metric tensor $g_{\mu\nu}$. Indeed,  it reads 	as 
\begin{equation}
{\cal L}=\frac{1}{2}g_{\mu\nu}\dfrac{dx^{\mu}}{d\sigma}\dfrac{dx^{\nu}}{d\sigma}
\label{bb2}
\end{equation}
where $\sigma$  represents  the affine parameter. The   metric form  gives 
\begin{equation}
{\cal L}=\frac{1}{2}
\left[
-f(r)\dot{t}^{2}
+\frac{\dot{r}^{2}}{f(r)}
+r^{2}\dot{\phi}^{2}
\right].
\label{bb3}
\end{equation}
In this scenario,    one has two conserved quantities being the energy $E$ and the angular momentum $L$  expressed as follows
\begin{equation}
E=f(r)\dfrac{dt}{d\sigma},
\qquad
L=r^{2}\dfrac{d\phi}{d\sigma}.
\end{equation}
The radial equation of motion of photons   is given by  
\begin{equation}
\left( \dfrac{dr}{d\sigma}\right) ^{2}+V_{\rm eff}=E^{2},
\label{bb6}
\end{equation}
where the effective potential  takes the following form 
\begin{equation}
V_{\rm eff}=\frac{L^{2}}{r^{2}}\,f(r).
\end{equation}
Using the metric function,   given in Eq. (\ref{ff}),  this scalar potential can be written as \begin{equation}
V_{\rm eff}
=
\frac{L^{2}}{r^{2}}
\left(
1-\frac{2M}{r}
-\frac{\Lambda}{3}r^{2}
+\frac{e^{-\gamma}Q^{2}}{r^{2}}
+m_{g}^{2}C
\left(
\frac{c_{1}r}{2}
+c_{2}C
\right)
\right)
\label{bb7}
\end{equation}
describing  the photon dynamics in  strong limits. In fact, it  depends on the black hole internal and external  parameters. Among others, it  includes  the ModMax parameter $\gamma$, and the massive gravity parameters $m_g$, $c_1$, $c_2$, and $C$. In this  way, the structure of photon trajectories and the existence of unstable circular orbits can be  determined in terms of such black hole  parameters. In Fig.~(\ref{fig:1}), we  show the effective potential  behaviors as a function of the radial coordinate.

These one dimensional  curves   illustrate the variation of  the charge parameter $Q$  for both weak and strong massive gravity models. One observes  that the maximum value of the effective potential $V_{eff}$  increases by increasing the black hole charge $Q$.

The  photon orbits are usually  circular and unstable, associated with   the maximum of the effective potential.  In this description,  the  radius of the photon sphere can be obtained  by imposing following constraints
\begin{equation}
V_{\mathrm{eff}}(r_p)=0,
\qquad
\left.
\frac{dV_{\mathrm{eff}}(r)}{dr}
\right|_{r=r_p}
=0,
\label{eq:12}
\end{equation}
where $r_p$  denotes   the photon sphere  radius. Indeed,  the constraint 
\begin{equation}
\left.
\frac{dr}{d\phi}
\right|_{r=r_p}
=0,
\label{eq:13}
\end{equation}
leads to   the following orbital equation
\begin{equation}
\frac{dr}{d\phi}
=
\pm r
\sqrt{
f(r)
\left(
\frac{r^{2}f(r_p)}
{r_p^{2}f(r)}
-1
\right)
}.
\label{eq:14}
\end{equation}
To get the desired equation, one  considers a light ray emitted by a stationary observer located at  $r_ 0$ propagating toward the past at an angle  $\beta$ relative to the radial direction. The  associated calculation   methods   have been explored in various   works  ~\cite{61,62,63}. Concretely,  the apparent shape of the black hole shadow can be elaborated using  the following   trigonometrical  relations
\begin{equation}
\cot\beta
=
\left.
\frac{\sqrt{g_{rr}}}
{\sqrt{g_{\phi\phi}}}
\frac{dr}{d\phi}
\right|_{r=r_0},\quad \sin^{2}\beta
=
\frac{f(r_0)\,r_p^{2}}
{r_0^{2}f(r_p)}.
\label{eq:15}
\end{equation}
Accordingly, the shadow radius observed at the position $r_0$  can take the form 
\begin{equation}
R_s
=
r_0\sin\beta
\label{eq:17}
\end{equation}
giving 
\begin{equation}
R_s
=
r_p
\sqrt{
\frac{f(r_0)}
{f(r_p)}
}.
\label{eq:18}
\end{equation}
Considering  Eqs.~(\ref{eq:12}) and (\ref{eq:14}), together with the condition
$dV_{\mathrm{eff}}/dr=0$, the photon sphere radius  can fulfill  the following algebraic equation 
\begin{equation}
8e^{-\gamma}Q^{2}
+r_p
\left[
-12M
+r_p
\left(
4
-
Cm_g^{2}
\left(
4Cc_{2}
-c_{1}r_p
\right)
\right)
\right]
=0.
\label{eq:19}
\end{equation}
It turns out that   an   analytical  solution  of such an equation  is a tough task. However,  one could follow a numerical method. In this way, one can obtain  the photon sphere radius. For a static observer located sufficiently far from the black hole, the corresponding shadow radius can then be  obtained using  Eq.~(\ref{eq:18}). To elaborate  the optical appearance of the black hole,  the celestial coordinates are needed  to  describe  the apparent position of photons on the observer sky.  Taking  an observer located far away from the black hole,  these  coordinates  are given by 
\begin{equation}
\begin{aligned}
X &=
\lim_{r\rightarrow\infty}
\left(
-r^{2}\sin\theta_{0}
\frac{d\phi}{dr}
\right)_{\theta_{0}=\pi/2},
\\
Y &=
\lim_{r\rightarrow\infty}
\left(
r^{2}
\frac{d\theta}{dr}
\right)_{\theta_{0}=\pi/2},
\end{aligned}
\end{equation}
where  the couple $(r_0,\theta_0)$ denotes   the observer position. It is  recalled that the coordinate $X$ measures the apparent horizontal displacement of the image.  The coordinate $Y$, however,   represents  the vertical displacement in the observer sky.

Now we are in a position  to  inspect  the  impact  of the black hole parameters on the geometric  behaviors  of   the shadows. To examine  the  influence of the electric charge on the shadows in both the weak and  the strong massive gravity regimes, we  use a numerical  method   to elaborate the configuration  contours by considering an appropriate   range for the black hole  parameter values.  This enables  the efficient determination of the shadow boundaries for different parameter configurations. For each simulation, all parameters are kept fixed except the electric charge $Q$, which varies  from one configuration to another with a step size of 0.01. 
\begin{figure}[th]
\centering
\begin{tabular}{cc}
\includegraphics[scale=0.4]{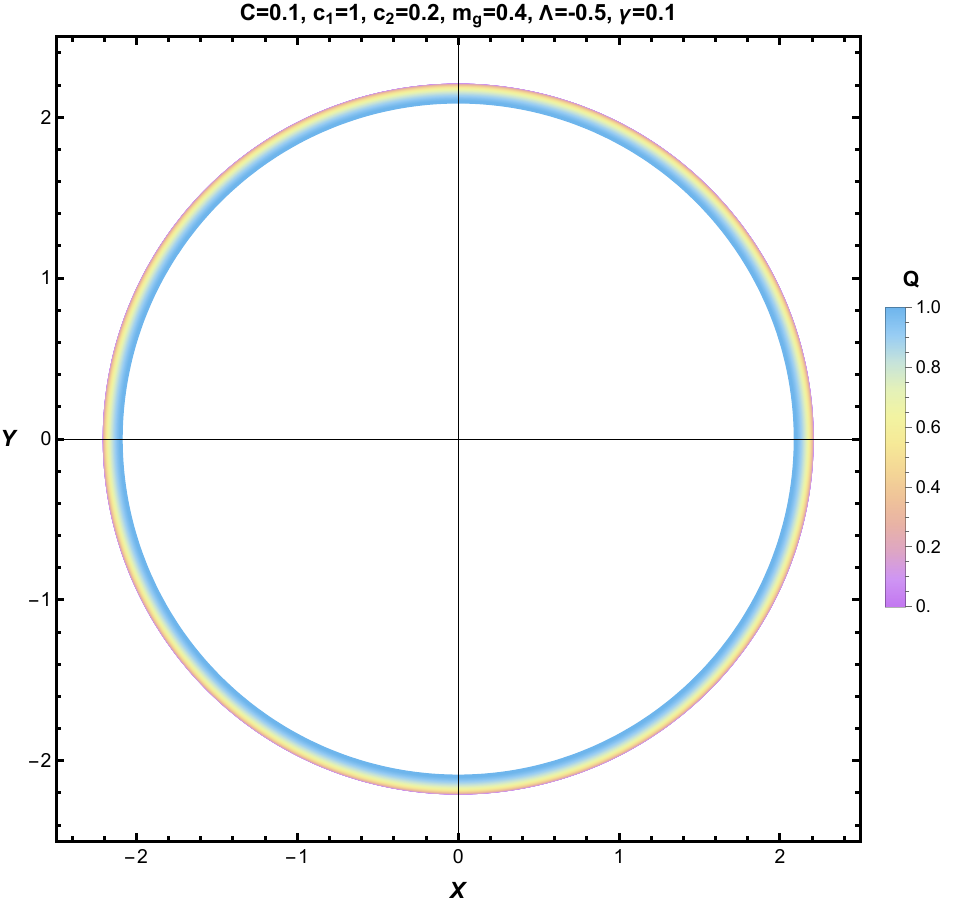}&
\includegraphics[scale=0.4]{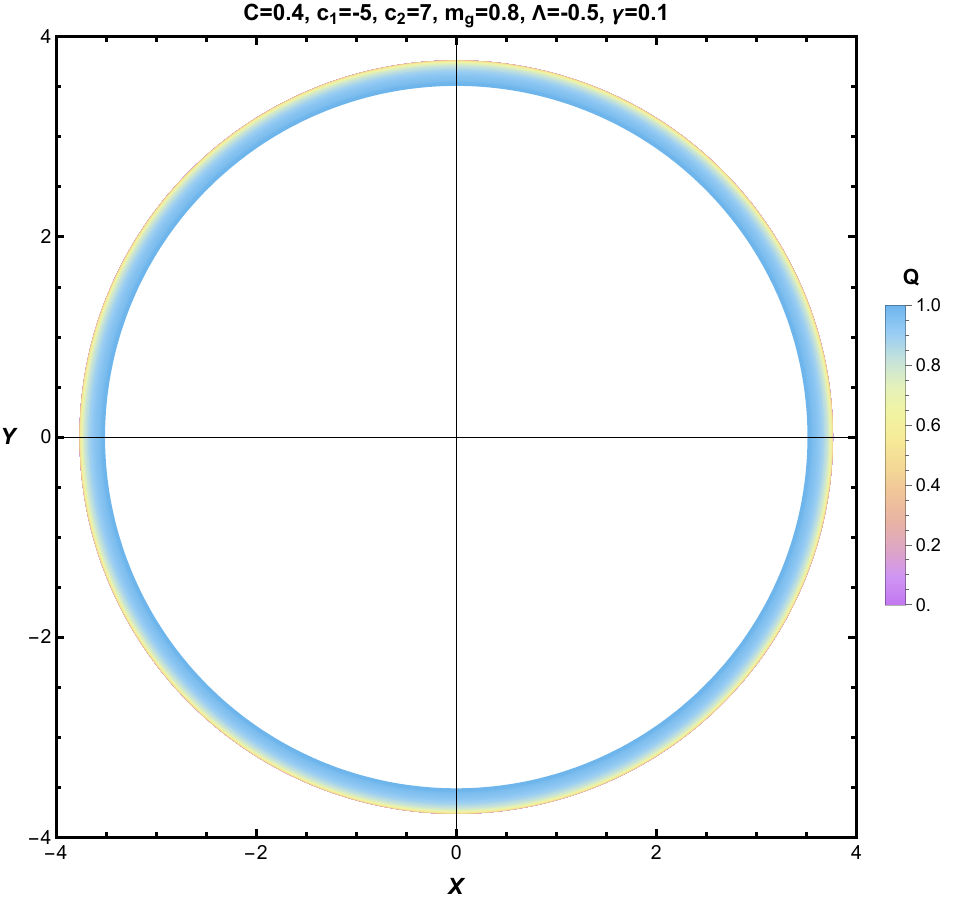}
\end{tabular}
\caption{Black hole shadow for different values of the physical parameters of the charged AdS black hole in dRGT-like massive gravity coupled to ModMax nonlinear electrodynamics.}
\label{sh1}
\end{figure}
Fig.~\ref{sh1}  shows   the dependence of the black hole shadows on the electric charge for both weak and strong massive gravities.  One observes that  the charge modifies the  size shadow radius. However, it   preserves its circular shape.  This is not surprising  since one considers   a  static and spherically symmetric spacetime without spin  effects. Fixing the values of the  electric charge, the weak massive gravity regime  produces  a smaller shadow radius than the strong massive gravity regime. These variations  generate  valuable insights into the optical properties of the black hole which can be exploited to provide observational signatures for testing modified theories of gravity with the help of  EHT  empirical findings.

\begin{figure}[ht!]
\centering
\begin{tabular}{cc}
\includegraphics[width=0.5\linewidth]{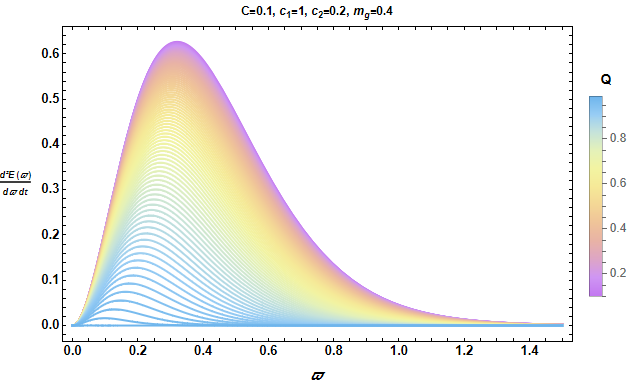} &
\includegraphics[width=0.5\linewidth]{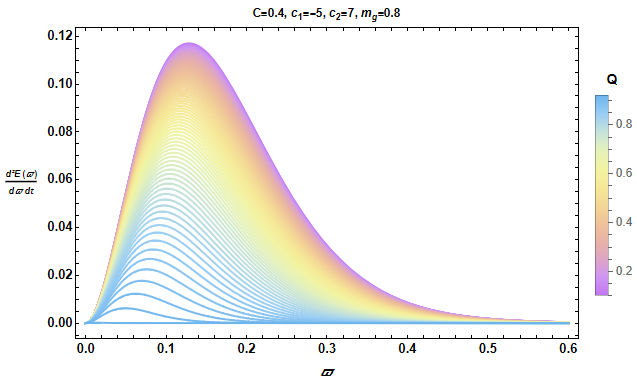}
\end{tabular}
\caption{Energy emission rate as a function of the emission frequency for different values of $Q$ with $\gamma=0.1$ and $\Lambda=-0.5$.}
\label{fig:77}
\end{figure}

A close inspection shows that  the photon sphere  can play a primordial  role in the    examination   of   the optical aspects  of the black hole spacetime. In the high-energy (geometrical optics) scenarios, the absorption cross section approaches a limiting constant value controlled by the unstable circular photon orbits. This  can be  related to the  shadow  of the black holes\cite{55,56}. Since the shadow identifies  the apparent optical size of the black hole,  the limiting absorption cross section can be determined by considering  the geometrical cross section passing through the relation
\begin{equation}
\sigma_{\rm lim}\simeq \pi R_s^2
\end{equation}
where  $R_s$  represents  the shadow radius. The associated  energy emission rate  takes  the form 
\begin{equation}
\frac{d^{2}E(\omega)}
{d\omega\,dt}
=
\frac{2\pi^{3}R_s^{2}\omega^{3}}
{\exp(\omega/T_H)-1},
\end{equation}
where $T_H$  represents  the Hawking temperature and $\omega$  denotes the emission frequency. This expression provides  a direct link between the thermodynamic properties  and the optical aspect of the black holes. This can  make  the energy emission spectrum a valuable observable to probe the geometric properties of the   spacetime. For the charged AdS black holes in dRGT-like massive gravity coupled to  the ModMax nonlinear electrodynamics, the Hawking temperature is  shown to be
\begin{equation}
T_H=
\frac{
3Cc_1m_g^2r_h^3
-24e^{-\gamma}Q^2
+24Mr_h
-8\Lambda r_h^4
}
{12\pi r_h^3}.
\label{TH}
\end{equation}
Since both the photon sphere radius and the Hawking temperature depend nonlinearly on the black hole parameters,  finding  analytical expressions for  the shadow radius and the corresponding energy emission rate is generally a  tough task.  This  can be determined  using numerical methods.  Indeed, we can approach  the relevant  quantities efficiently  by considering acceptable regions of the black hole  moduli space. Fig.~(\ref{fig:77})   illustrates  the behavior of the energy emission rate as a function of the emission frequency for different values of  the electric charge $Q$. 

 Increasing the charge,  
the energy emission rate decreases.  Moreover,  one observes that   the weak massive gravity exhibits a higher maximum peak than the  strong massive gravity. The energy emission rate can supply  an additional observable  complementing the shadow analysis.  This can be used  for  distinguishing such  black hole solutions from alternative gravitational models using  observational signatures in astrophysics.

\subsection{Constraining black hole parameters from EHT observations}
In this subsection, we would like  to establish a connection between the obtained theoretical predictions and  the observational data   with the help of  the empirical   results reported by the EHT international collaboration.  More specifically,   we  constrain the parameters of such  black hole models, by   exploiting observational data from the black holes  of   M87* and Sgr A*  types ~\cite{19,20, 2000, AE1,AE2,AE3}.  In this way,  the constrained values can be  determined  using   the difference in the diameter of the black hole shadow relative to that of a Schwarzschild black hole. Indeed, it is expressed as 
\begin{equation}
{\ d} = \frac{R_s}{r_{sh}}-1,
\end{equation}
where $R_s$  denotes  the shadow radius and $r_{\text{sh}}$  represents the Schwarzschild radius. In order  to  confront the  rational models with empirical  data,     the dimensionless ratio $R_s/M$  can be considered  as a primordial observable. The  $1-\sigma$  and  $2-\sigma$  confidence intervals derived from the EHT observations are displayed in Table ~\ref{t1}.

\begin{figure}[h!]
\centering
\includegraphics[scale=0.3]{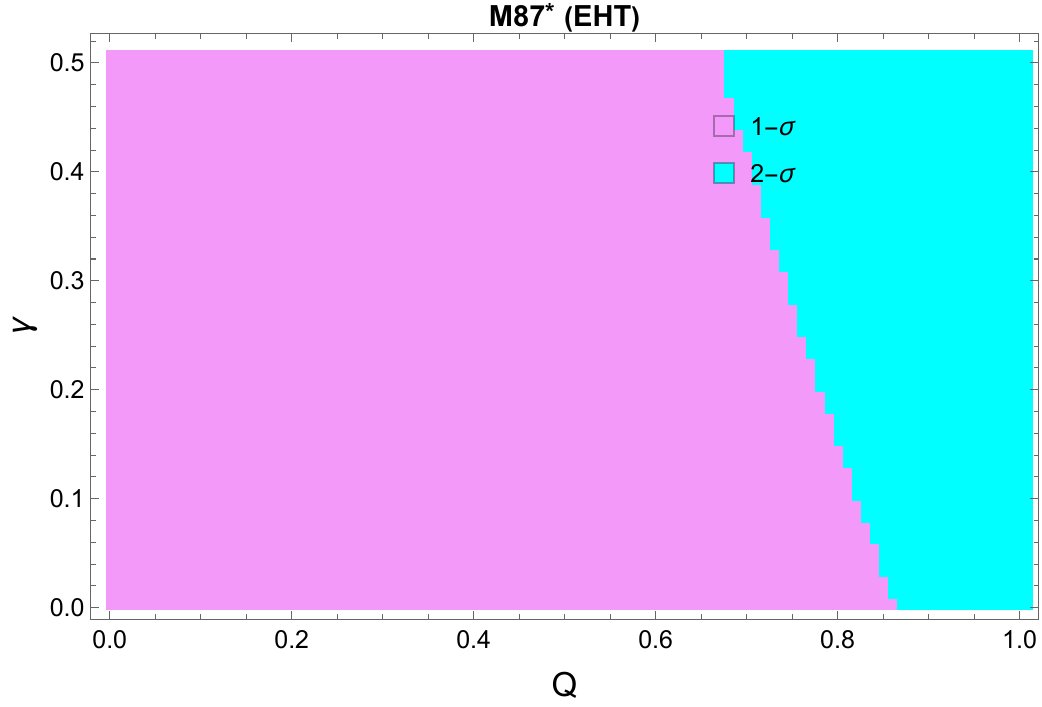} \includegraphics[scale=0.3]{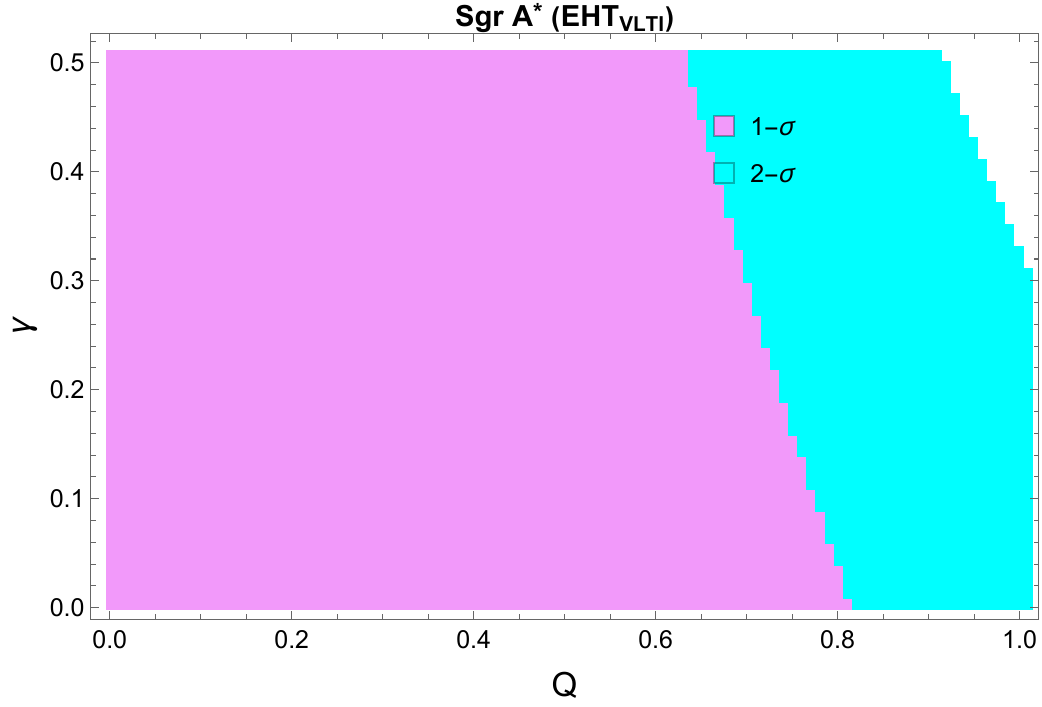}
\includegraphics[scale=0.3]{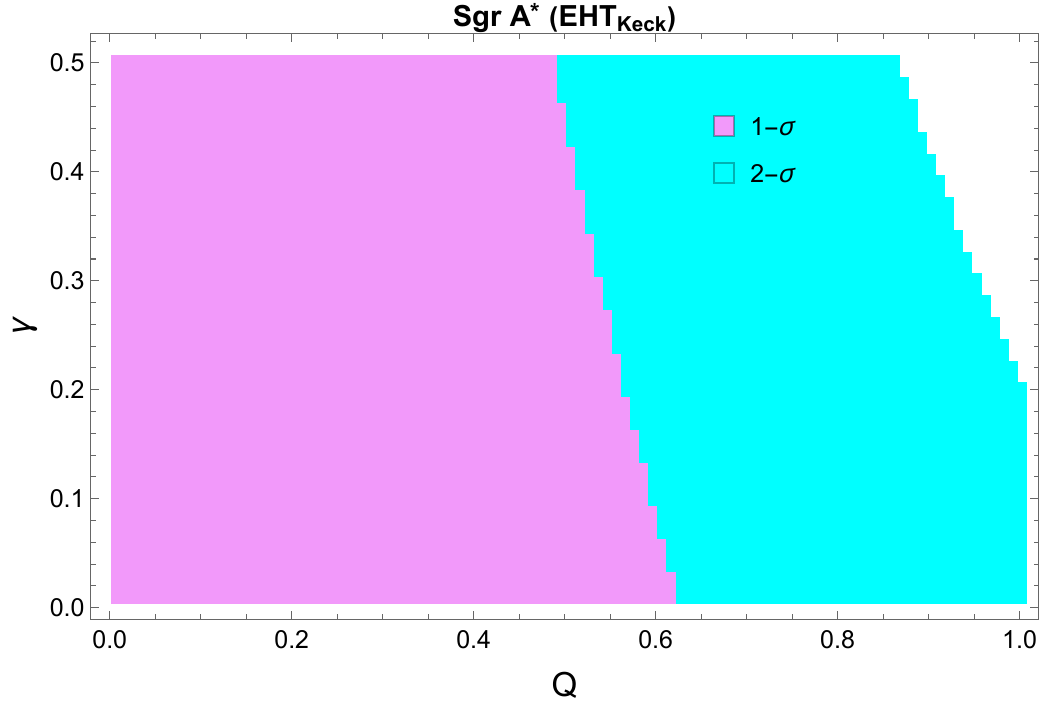}
\includegraphics[scale=0.3]{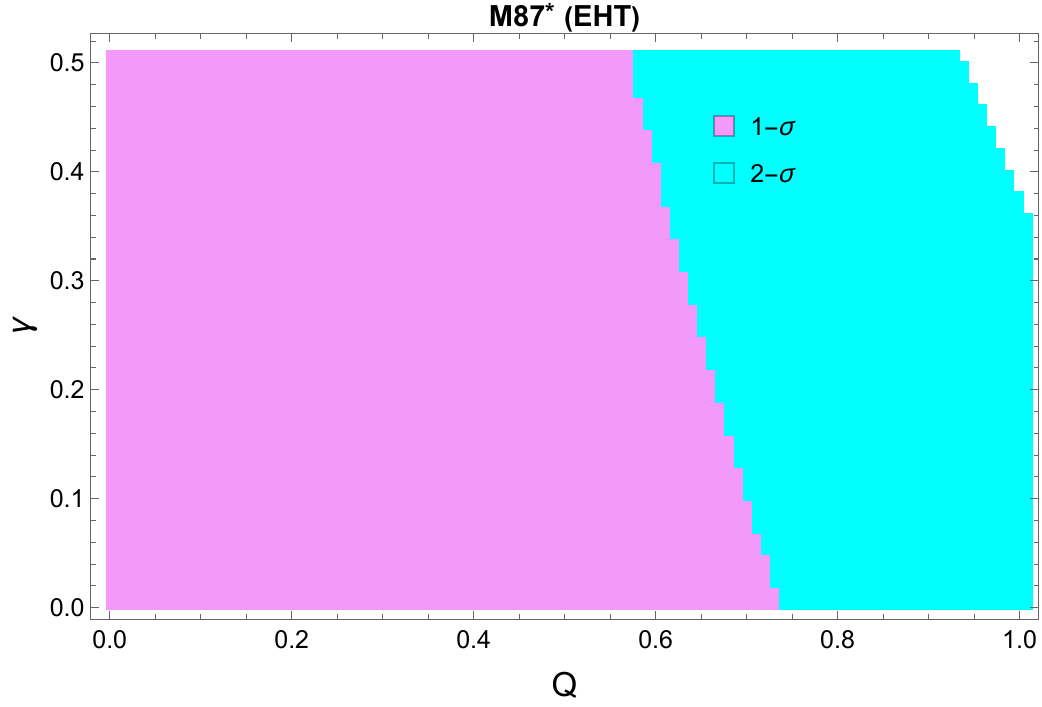} \includegraphics[scale=0.3]{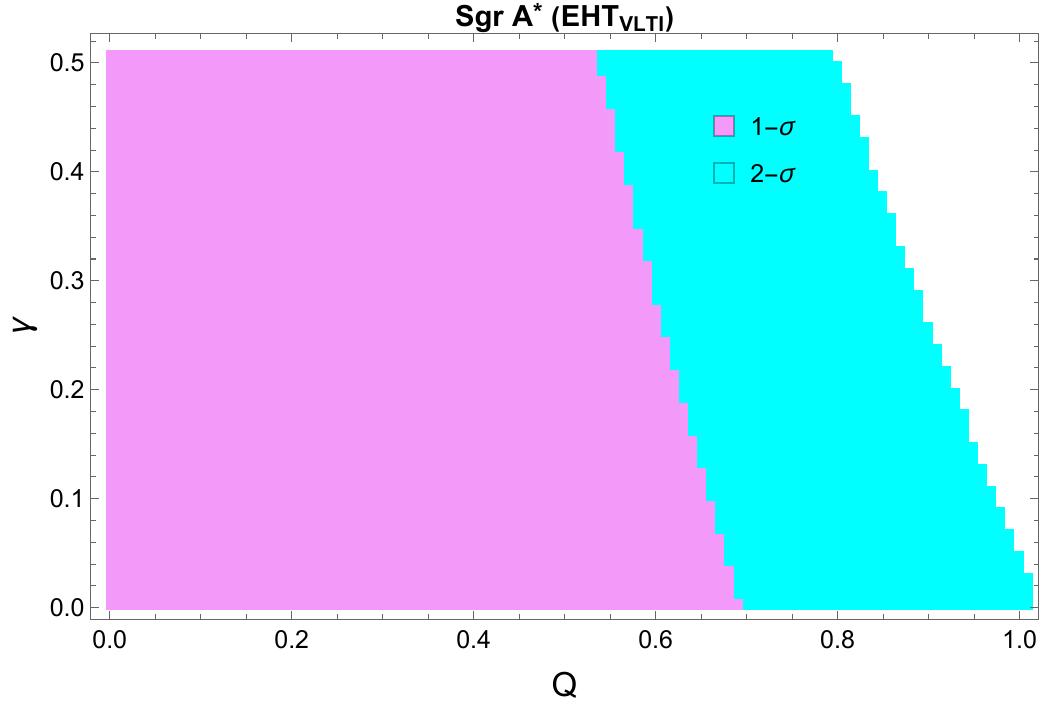}
\includegraphics[scale=0.3]{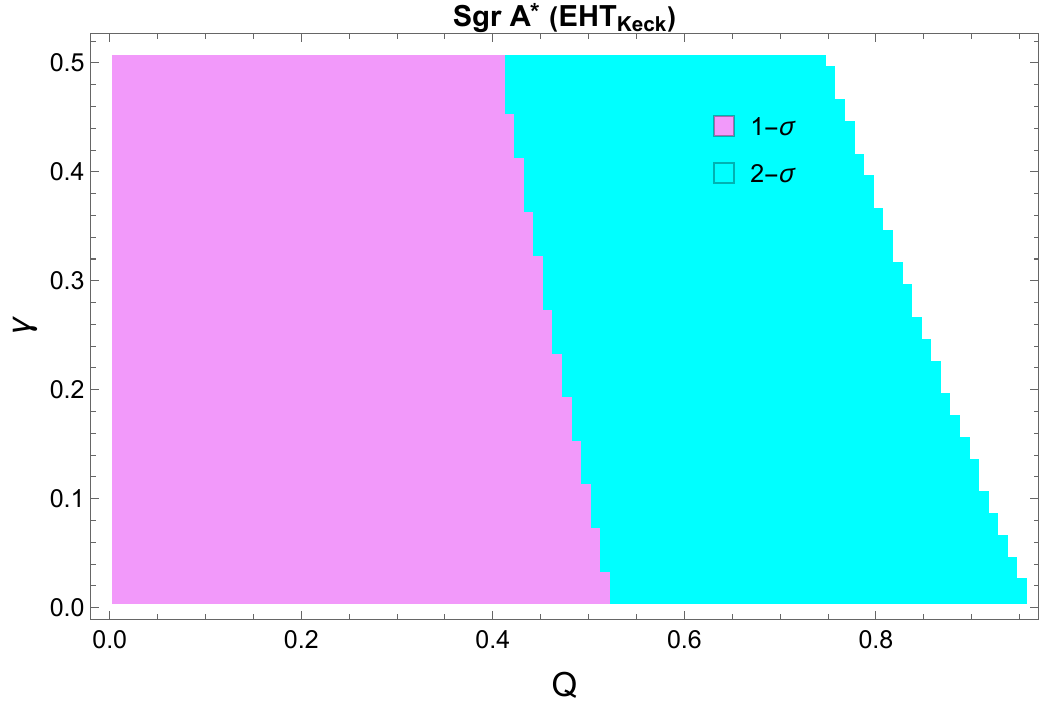}
\caption{\textit{\protect\footnotesize Constraint regions in the $(\gamma, Q)$ plane for weak massive gravity (top row) and strong massive gravity (bottom row).}}
\label{ss1}
\end{figure}

\begin{table}[h!]
\centering
\begin{tabular}{|c|c|c|c|}
\hline
{Black hole} & {Deviation ($d$)} & {1-$\sigma$ bounds}
& {2-$\sigma$ bounds} \\ \hline
M87$^*$ (EHT) & $-0.01^{+0.17}_{-0.17}$ & $4.26 \leq \frac{R_s}{M} \leq 6.03$
& $3.38 \leq \frac{R_s}{M} \leq 6.91$ \\ \hline
Sgr~A$^*$ (EHT$_{\text{VLTI}}$) & $-0.08^{+0.09}_{-0.09}$ & $4.31 \leq \frac{%
R_s}{M} \leq 5.25$ & $3.85 \leq \frac{R_s}{M} \leq 5.72$ \\ \hline
Sgr~A$^*$ (EHT$_{\text{Keck}}$) & $-0.04^{+0.09}_{-0.10}$ & $4.47 \leq \frac{%
R_s}{M} \leq 5.46$ & $3.95 \leq \frac{R_s}{M} \leq 5.92$ \\ \hline
\end{tabular}%
\caption{ \textit{\protect\footnotesize  Fractional deviations and
corresponding bounds for M87$^*$ and Sgr~A$^*$ black holes.}}
\label{t1}
\end{table}
 In this  analysis, we consider a fixed value of  the black hole mass being $M=1$. 
To do so,  we  use parallel programing computations to  determine the reduced moduli space   pairs  $(\gamma, Q)$   providing the  black hole shadow configurations being  consistent with the EHT observational constraints within the $1-\sigma$ and $2-\sigma$ confidence intervals. The obtained results for the  weak and  the strong massive gravity models  are  shown  in Fig.~(\ref{ss1}).

It has been shown that  the density of the parameter points being  consistent with the observational data decreases as the electric charge $Q$ increases. This behavior shows   that the spacetime of the underlying black hole  reproduces the observed shadow characteristics over a more limited region of  the  reduced moduli  space. Since the parameters are  correlated with each other, we constrain them by fixing one while varying the other.  Table $\ref{tab11}$, for instance,   illustrates  the appropriate  ranges of $Q$ in the weak and strong gravity regimes, being  consistent with the $1-\sigma$ and $2-\sigma$ confidence intervals derived from the three black hole observations.
\begin{table}[!ht]
\centering
\caption{Constraints on the parameter $Q$ for weak and strong massive gravity, derived from observational bounds at the $1-\sigma$ and $2-\sigma$ confidence levels.}
\label{tab11}
\begin{tabular}{llccc}
\toprule
{Massive gravity} & {Confidence} & {M87$^*$} 
& {SgrA$^*_{\mathrm{Keck}}$} & {SgrA$^*_{\mathrm{VLTI}}$} \\
\midrule

\multirow{2}{*}{Weak} 
& $1-\sigma$ & $0.01 < Q < 0.8$ & $0.01 < Q < 0.75$ & $0.01 < Q < 0.57$ \\
& $2-\sigma$ & $0.01 < Q < 1$ & $0.01 < Q < 1$ & $0.01 < Q < 1$ \\
\midrule

\multirow{2}{*}{Strong} 
& $1-\sigma$ & $0.01 < Q < 0.68$ & $0.01 < Q < 0.64$ & $0.01 < Q < 0.48$ \\
& $2-\sigma$ & $0.01 < Q < 1$ & $0.01 < Q < 0.91$ & $0.01 < Q < 0.86$ \\

\bottomrule
\end{tabular}
\end{table}
A thorough examination of the results obtained in weak and strong mass gravity regimes reveals that the observational constraints become increasingly rigorous as the magnitude of the mass gravity increases.  More specifically, the weak  massive gravity regime features significantly larger regions in the reduced module space $(\gamma,Q)$   that remain consistent with EHT observations.  The strong massive gravity regime, however,  significantly decreases such  acceptable  regions. Moreover, the electric charge $Q$  can be considered  as a relevant quantity in the constraining parameter scenarios. Indeed, the  number of admissible solutions systematically decreases  by increasing the electric  charge   $Q$. These results indicate that the  EHT shadow measurements  provide   strict constraints on the electric charge $Q$. Moreover, they reveal  that  the stronger massive gravitational corrections  reduce the range of viable black hole parameters.

\section{Learning    black hole parameters with neural networks   through  EHT observations}
In this section, we present a  machine learning  scenario  with NNs to   approach black hole parameters via shadow  computations. This approach is motivated by the growing success of machine learning in gravitational physics, particularly in the study of black holes and non-trivial theories such as  string theory compactifications on Calabi--Yau  geometries. More recently, machine learning algorithms have been successfully applied to several problems in  black hole physics. This  includes the classification of black holes in the mass--spin parameter space and the investigation of quantum gravitational phenomena  and  the study  of Schwarzschild geodesic motion using   Lagrangian neural networks\cite{E34,E35,E355,E3550}. Likewise, in the context of string theory compactifications, such  network models have been utilized  to examine and classify complex geometric structures, such as  complete intersection Calabi--Yau manifolds and their freely acting quotients  in orbifold constructions\cite{E36}.  It has been suggested   that machine learning can efficiently identify physically relevant regions of  the moduli  spaces  distinguishing different classes of black holes. Moreover, they predict stability domains, and reveal hidden correlations being  often tough to extract using conventional numerical approaches.  Supported  by  such  physical   studies, we use a feed-forward fully connected neural network (FCNN) \cite{E37} to classify black hole shadow configurations according to their compatibility with the observational constraints reported by  EHT   international collaborations.

Among the various supervised learning algorithms available, the FCNN is adopted thanks to its  simplicity, computational efficiency, and ability to learn nonlinear relationships between the physical parameters and the corresponding observational constraints \cite{ML1,ML2}. It has been observed that   this  approach is particularly well  adopted  for  certain issues having  a  small number of input variables.   It  could provide stable predictions with low computational costs. After being trained, the FCNN functions are considered  as  computational approximation models  being able to identify regions in the reduced black hole module space that are consistent with EHT observations. This could be done by  evincing  the need to repeatedly perform highly complex numerical simulations of the black hole  shadows.

\subsection{Data preparation}
We start by discussing the  data preparation   used in  this work.  Indeed, 
 the datasets are generated via numerical codes   relaying on   parallel  programming   computations.  For each pint the reduced moduli  $(\gamma,Q)$, the associated  black hole shadow is numerically approached  and compared with the observational findings  reported by the EHT international  collaborations. This procedure is repeated for  different observational conditions,  providing  distinct datasets corresponding to the M87$^*$, Sgr A$^*_{\mathrm{VLTI}}$, and Sgr A$^*_{\mathrm{Keck}}$ observations,  where each one is  considered at the $1\sigma$ and $2\sigma$ confidence levels.  Based on such a  comparison, each sample is attributed a binary label.  In this way,  a value of 1  denotes  that the computed shadow satisfies the corresponding EHT observational constraints. However, a value of 0   corresponds  to an incompatible configuration. In this way, one has the following scheme
\begin{equation}
(\gamma,Q)\equiv  0,1.
\end{equation}
To guarantee the quality of the dataset, a filtering procedure  has been  implemented to eliminate non-physical and redundant configurations, keeping  only parameter values  falling  within physically allowed ranges and  matching with  the results of the EHT collaboration. The resulting dataset is then balanced by selecting an equal number of compatible and incompatible configurations before being randomly  intermixed. In order to improve the convergence and  the stability of the training process, the input features are subsequently standardized using the \texttt{StandardScaler} transformation. This procedure  yields  zero mean and unit variance. Then, the datasets are split into independent training, validation, and test subsets by using  a fixed random seed (\texttt{random\_state = 42}).  These data  scenarios  guarantee the  reproducibility. Moreover, they  prevent   information leakage between the different subsets.  It turns out that  this  approach provides  a clean, balanced, and well-structured dataset suitable for a supervised binary classification. In  fact, this  enables  the FCNN to efficiently distinguish between black hole parameter configurations being compatible or incompatible with the EHT observational  findings.
It should be noted in passing that the machine learning techniques presented in this work are not intended to directly infer black hole parameters from EHT observational images, nor to replicate the full Bayesian estimation of parameters carried out by the international EHT collaborations.  Instead, the FCNN  has  exploited  as a computational surrogate model trained in a supervised manner to efficiently identify the  parameter space regions   matching with EHT observational findings.  In this way,  the  binary labels are generated by comparing the numerically computed the  black hole shadows with the corresponding EHT observational confidence intervals. Once trained, the FCNN rapidly predicts whether a given parameter space  configuration is compatible with such EHT  empirical constraints. To do so, it is crucial  to avoid repeated shadow calculations during  he  explorations of the  black hole parameter space. An  evaluation of the probability, including  the noise parameters, the covariance matrices, and  the uncertainties related to the emission model, is beyond the scope of this study. It would be the subject of future works.

To train the classifier, we use  a feedforward FCNN. This  takes the input tuple $(\gamma,Q)$ and predicts the associated binary label (0 or 1). To improve the robustness of the predictions and reduce the sensitivity of the classifier to small numerical variations, we adopt a Monte Carlo voting scenario. For each normalized input sample, 20 perturbed realizations have been  produced  by adding Gaussian noises with a standard deviation of 0.01 to the input features. Each perturbed realization is independently processed by  such a  FCNN. Then,  the final prediction can be  obtained  via   a majority voting over such 20 individual predictions.  In this way, the Gaussian perturbations are implemented  in order to to examine  the robustness of the trained classifier with respect to  small numerical variations of the normalized input features. The compatibility labels are not recalculated for perturbed  samples, since the primary goal is to examine  the stability of the classifier rather than to redefine the physical class of each instance. Indeed,  each normalized input parameter set is transformed into 20 Gaussian-perturbed realizations.  Moreover, each realization is independently processed by the FCNN. In this way,  the resulting predictions are combined through a  majority voting to produce the final output label. In fact, this procedure indicates  whether the corresponding black hole shadow matches with the EHT observational constraints.  This strategy not only  enriches the robustness of the classifier against small numerical perturbations but also preserves  the physical interpretation of the input parameters. Combined with the data preparation procedure, it  produces    stable, reliable, and accurate classification frameworks.

\begin{figure}[!h]
\centering
\includegraphics[scale=0.28]{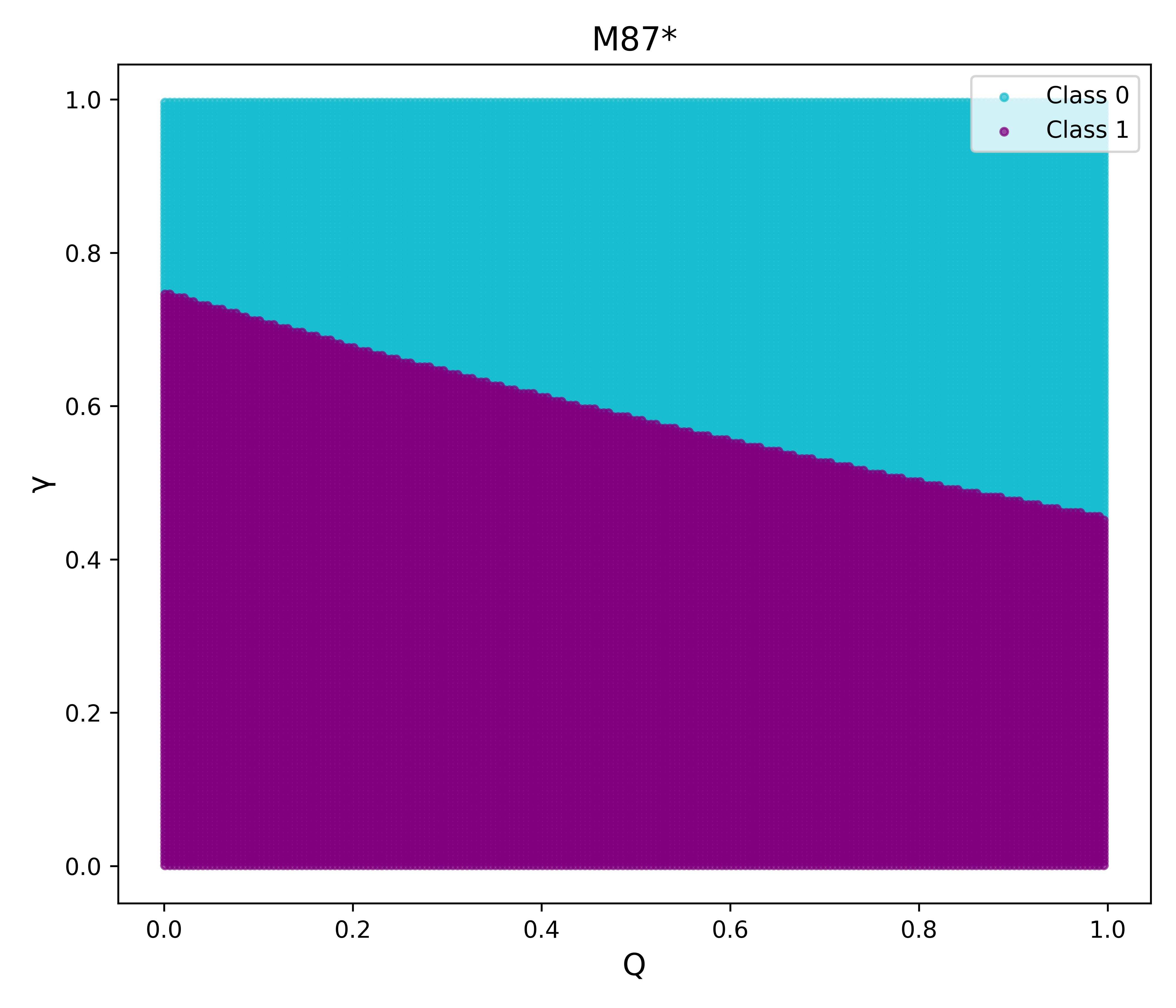}\hspace{2mm}
\includegraphics[scale=0.28]{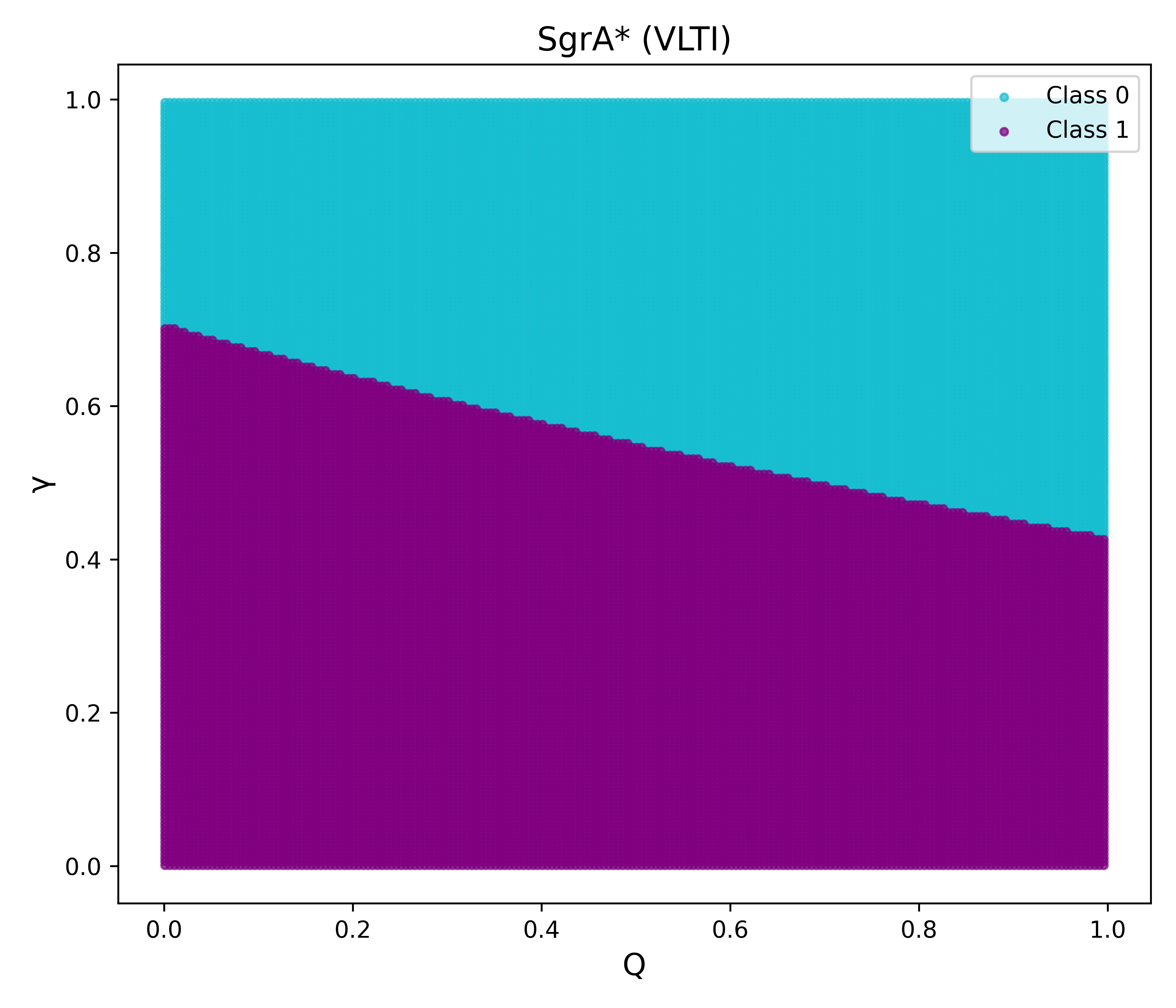}\hspace{2mm}
\includegraphics[scale=0.28]{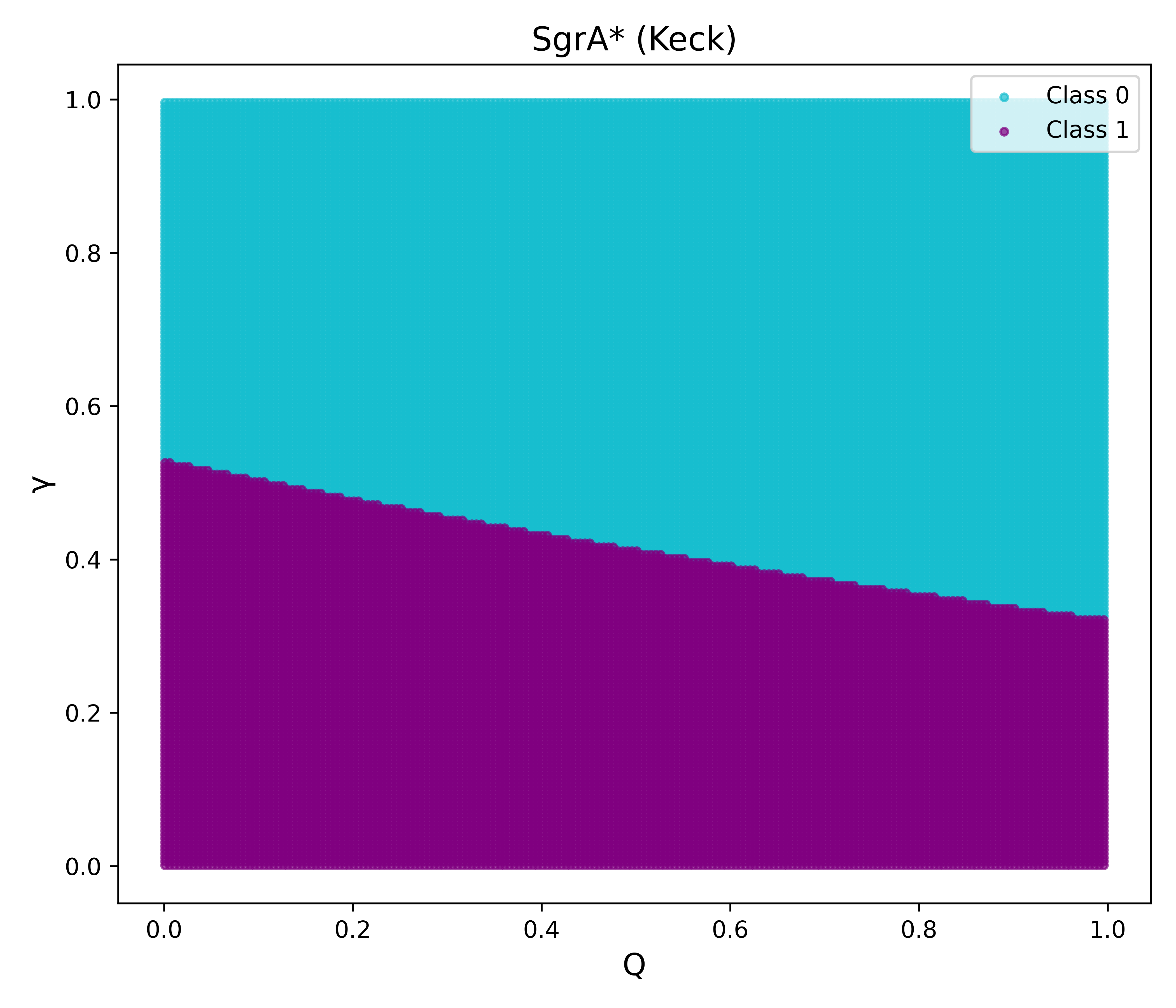}
\includegraphics[scale=0.28]{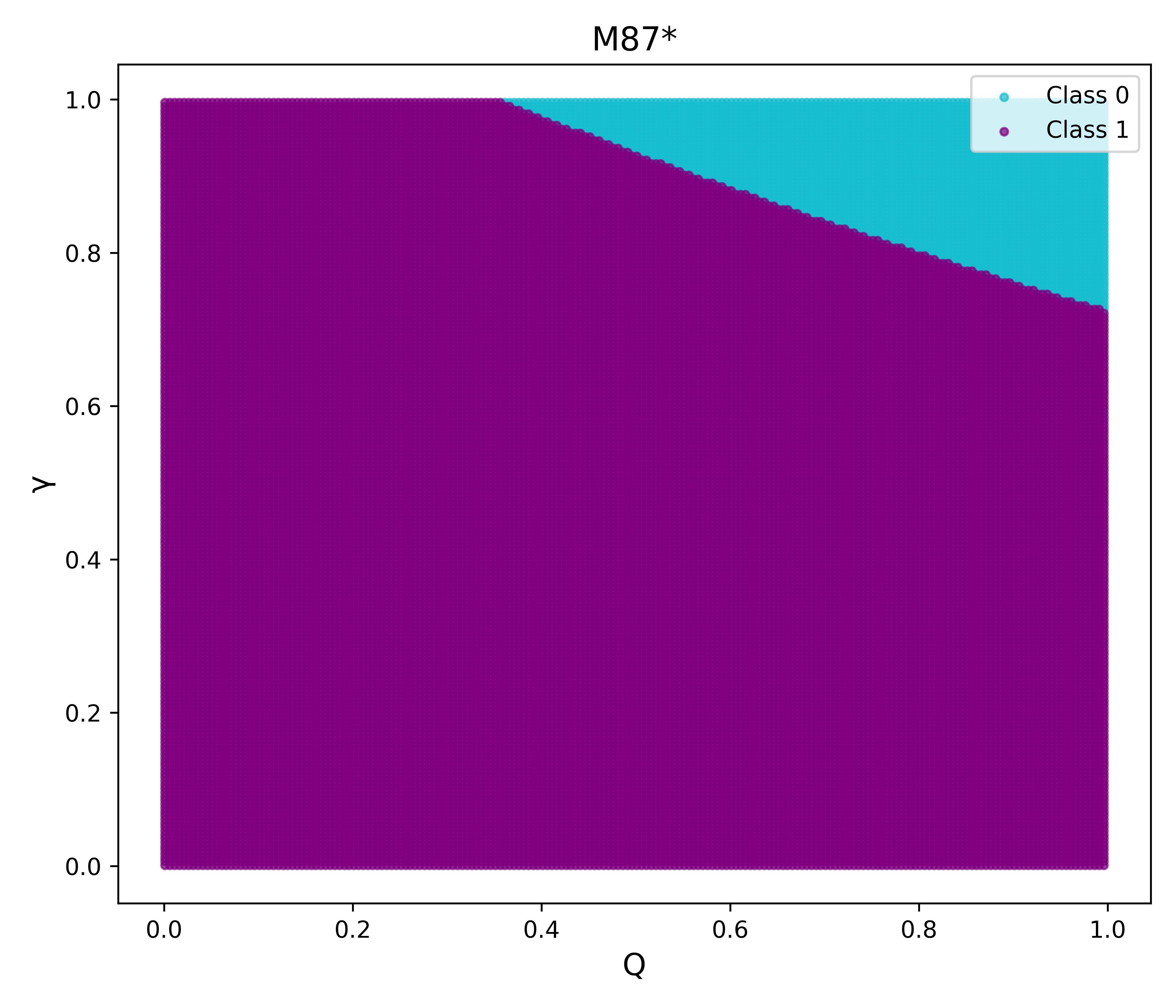}\hspace{2mm}
\includegraphics[scale=0.28]{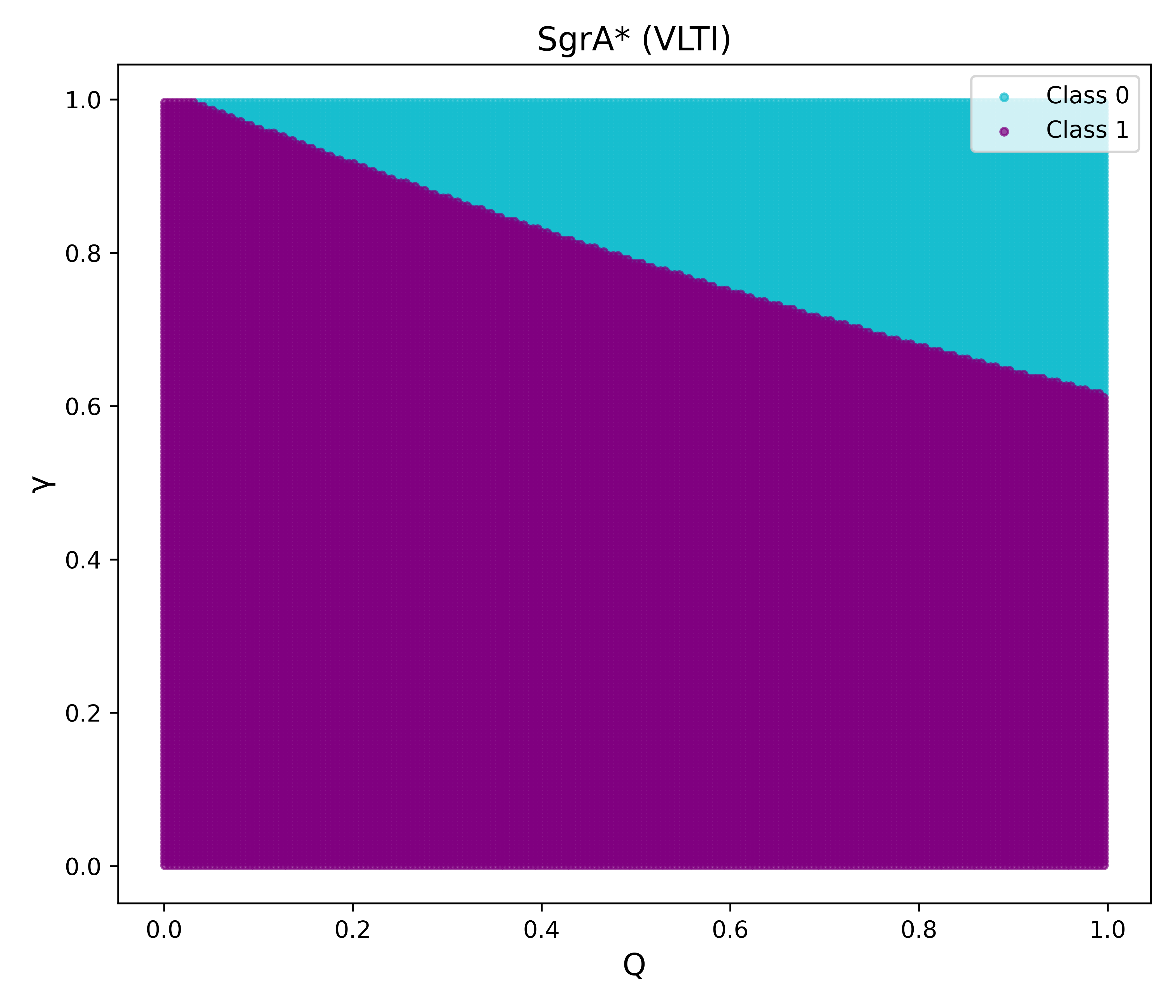}\hspace{2mm}
\includegraphics[scale=0.28]{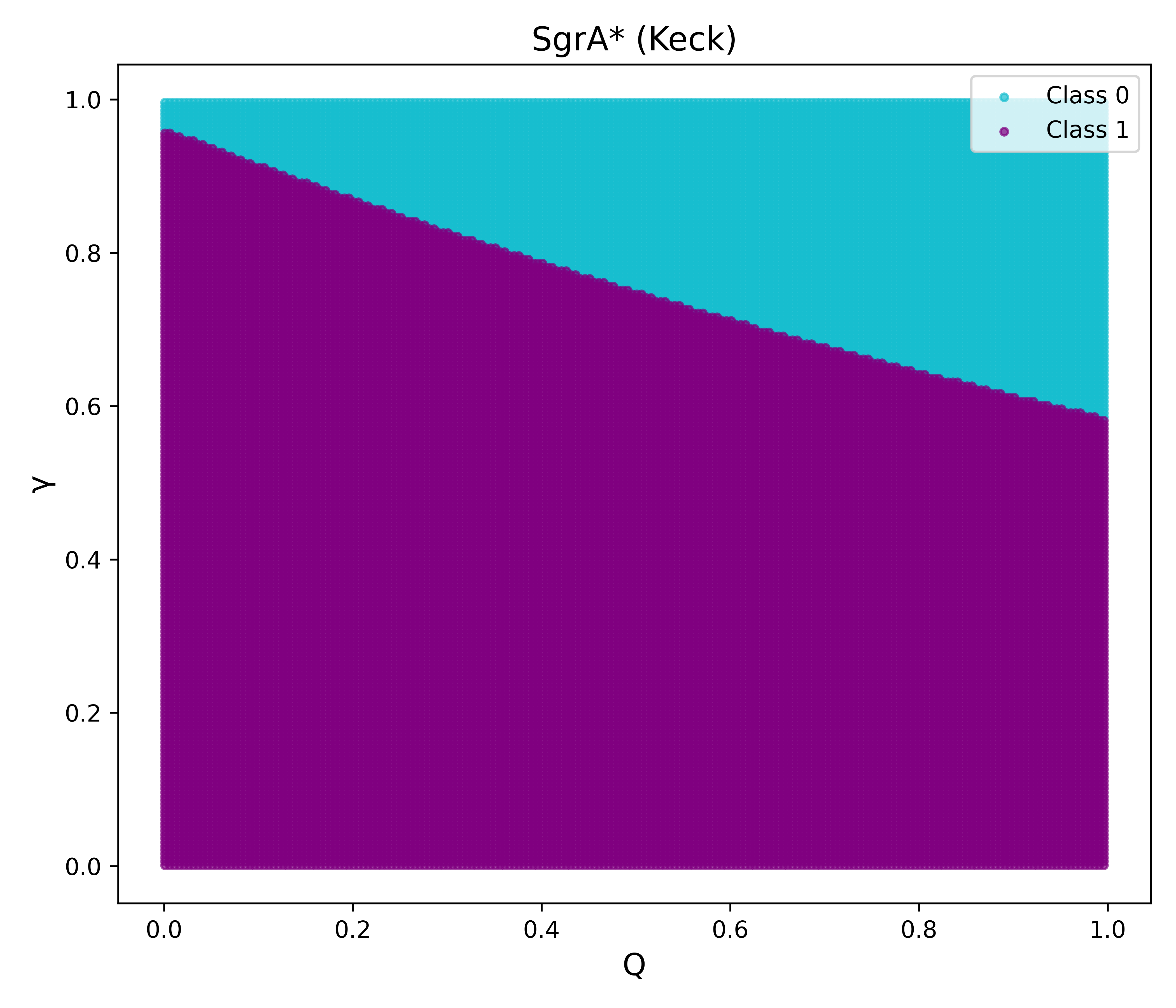}
\caption{Distributions of the generated datasets for the M87$^*$, Sgr A$^*_{\mathrm{VLTI}}$, and Sgr A$^*_{\mathrm{Keck}}$ observational datasets under the $1$--$\sigma$ (top row) and $2$--$\sigma$ (bottom row) confidence levels.}
\label{fig:train12}
\end{figure}

To  approach the generated datasets, Fig.~(\ref{fig:train12}) illustrates  the distributions of the compatible (Class~1) and the  incompatible (Class~0) black hole configurations for the M87$^*$, Sgr A$^*_{\mathrm{VLTI}}$, and Sgr A$^*_{\mathrm{Keck}}$ observational datasets under both the $1$--$\sigma$ and $2$--$\sigma$ confidence levels.  
Each point  in such regions of the moduli space  generates a  black hole configuration.  In this way,  the   binary label is assigned according to whether the corresponding black hole shadow matches with  the EHT observational constraints.  It  follows from this figure  that the compatible and  the incompatible configurations produce  well-defined regions separated by smooth nonlinear boundaries for all observational datasets.  Although the shape and the  location of these boundaries vary from one dataset to another, they consistently  exhibit  a continuous dependence of  the shadow compatibility on the involved   black hole  parameters. The limited overlap between the two classes  reveals  that the generated datasets are well-structured and suitable for   learning procedures. Furthermore, the gradual transition between compatible and incompatible regions  shows  that the decision boundary is controlled by nonlinear correlations among the physical parameters rather than by simple threshold values. Such a  organized  distribution  provides  favorable conditions for the FCNN to accurately learn the underlying decision boundary. Effectively,  it can   identify parameter  ranges  matching with  EHT observational constraints.

\subsection{FCNN classification of   shadow consistency  via  EHT observations}
Here, we approach the consistency of the obtained  black hole shadow configurations with the observational constraints generated by the EHT international collaborations using a FCNN scenario. In this way,   the  dataset consists of $40000$ samples, given in terms of the physical parameters $(\gamma,Q)$. Before splitting the dataset, an equal number of configurations compatible and incompatible with the EHT observational constraints is selected in order to ensure  a class balance.  The obtained  dataset is randomly  sampled and divided into training and test subsets using a fixed random seed (\texttt{random\_state = 42})  to ensure reproducibility.  During training, 10\% of the training subset is further reserved for a validation procedure.

\begin{figure}[!h]
\centering
\includegraphics[scale=0.32]{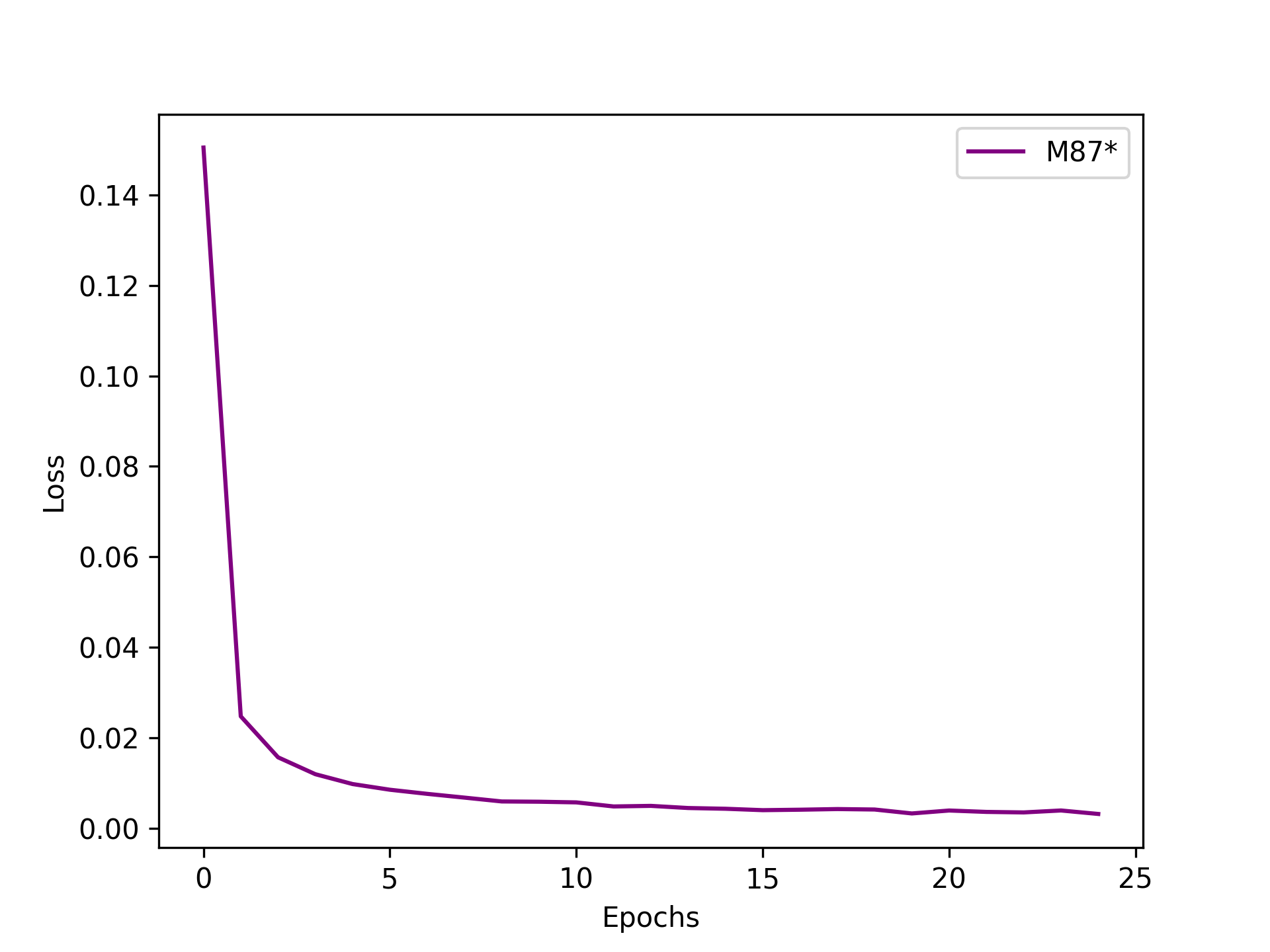}\hspace{2mm}
\includegraphics[scale=0.32]{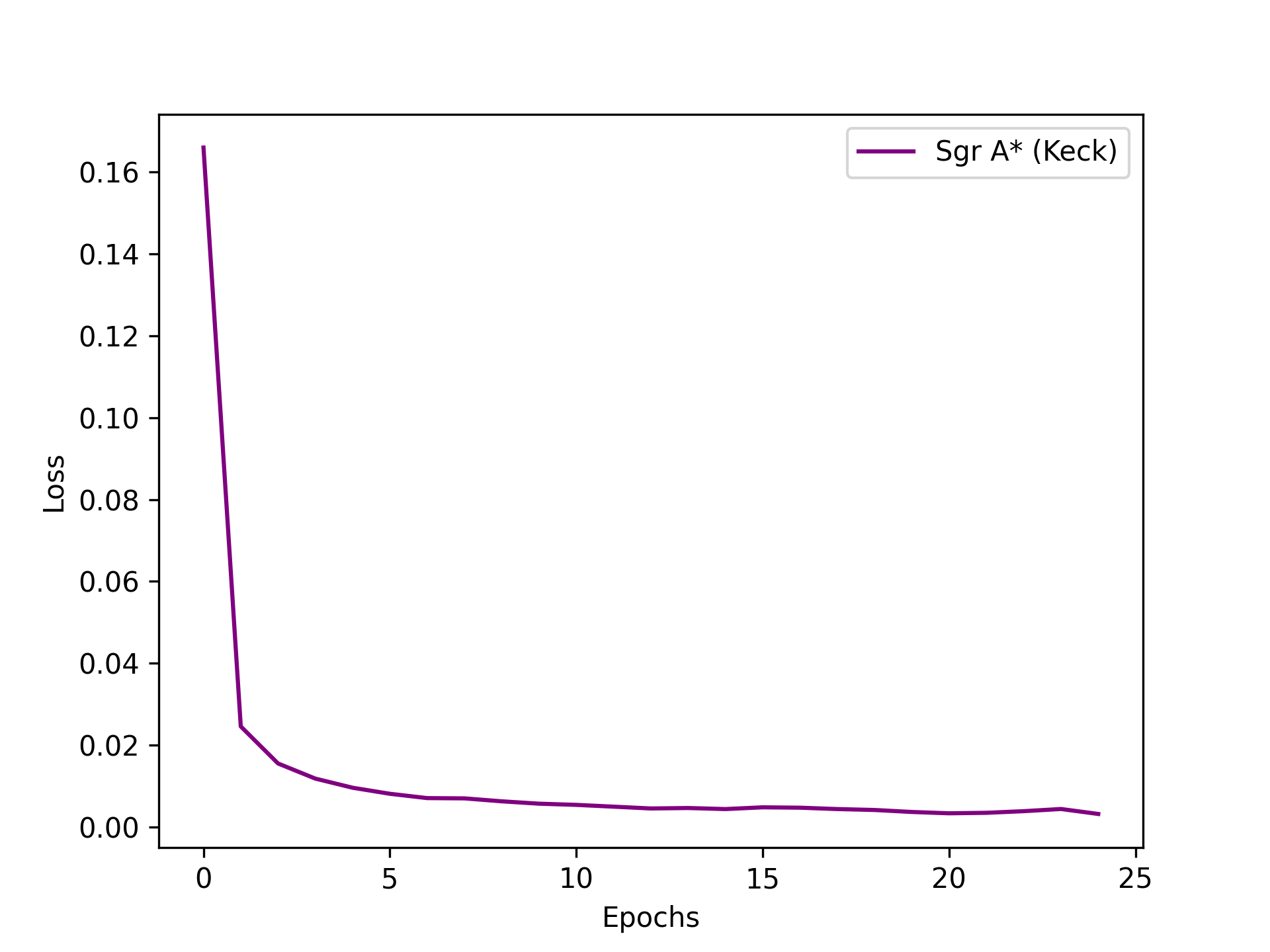}\hspace{2mm}
\includegraphics[scale=0.32]{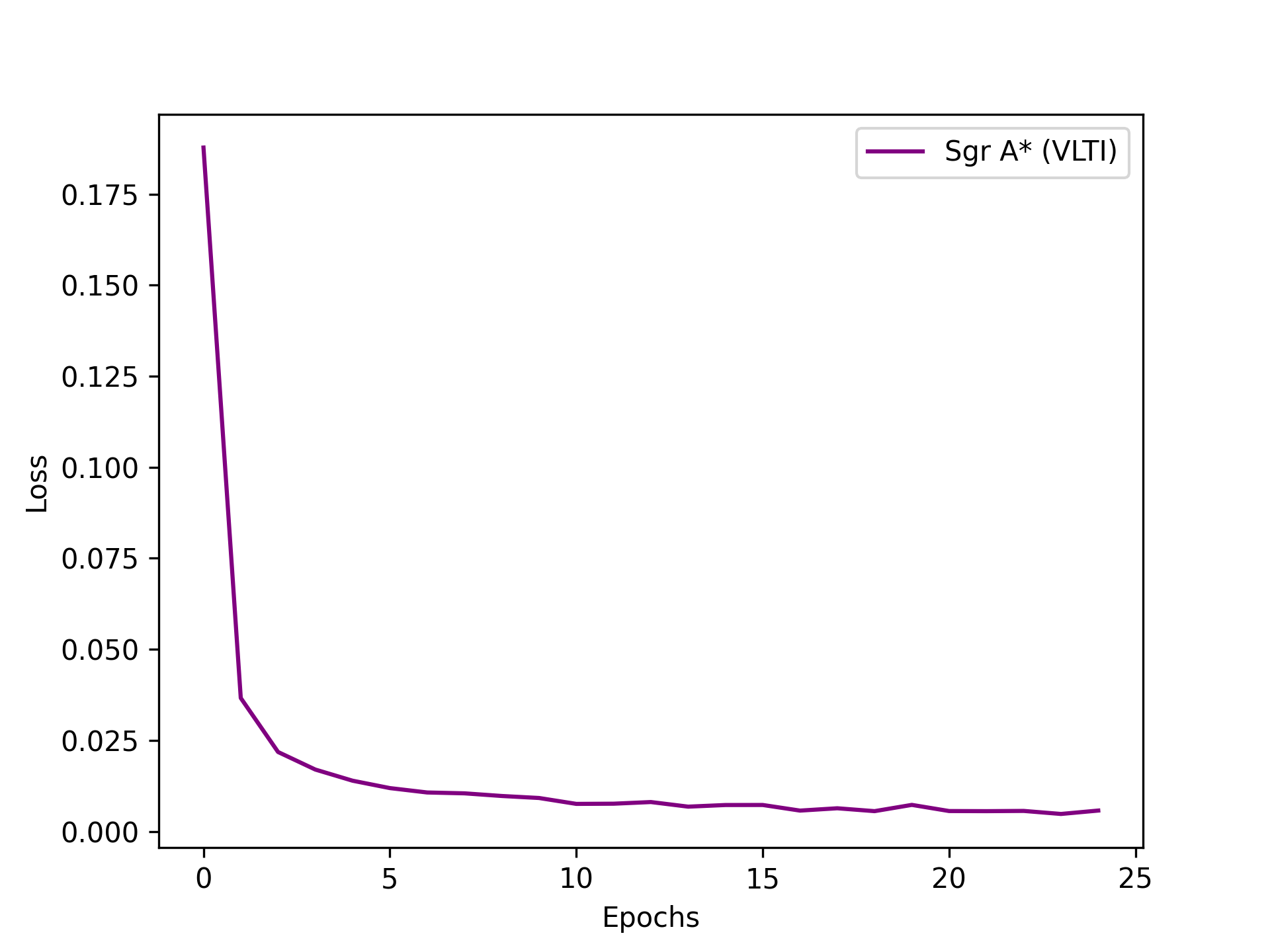}
\includegraphics[scale=0.32]{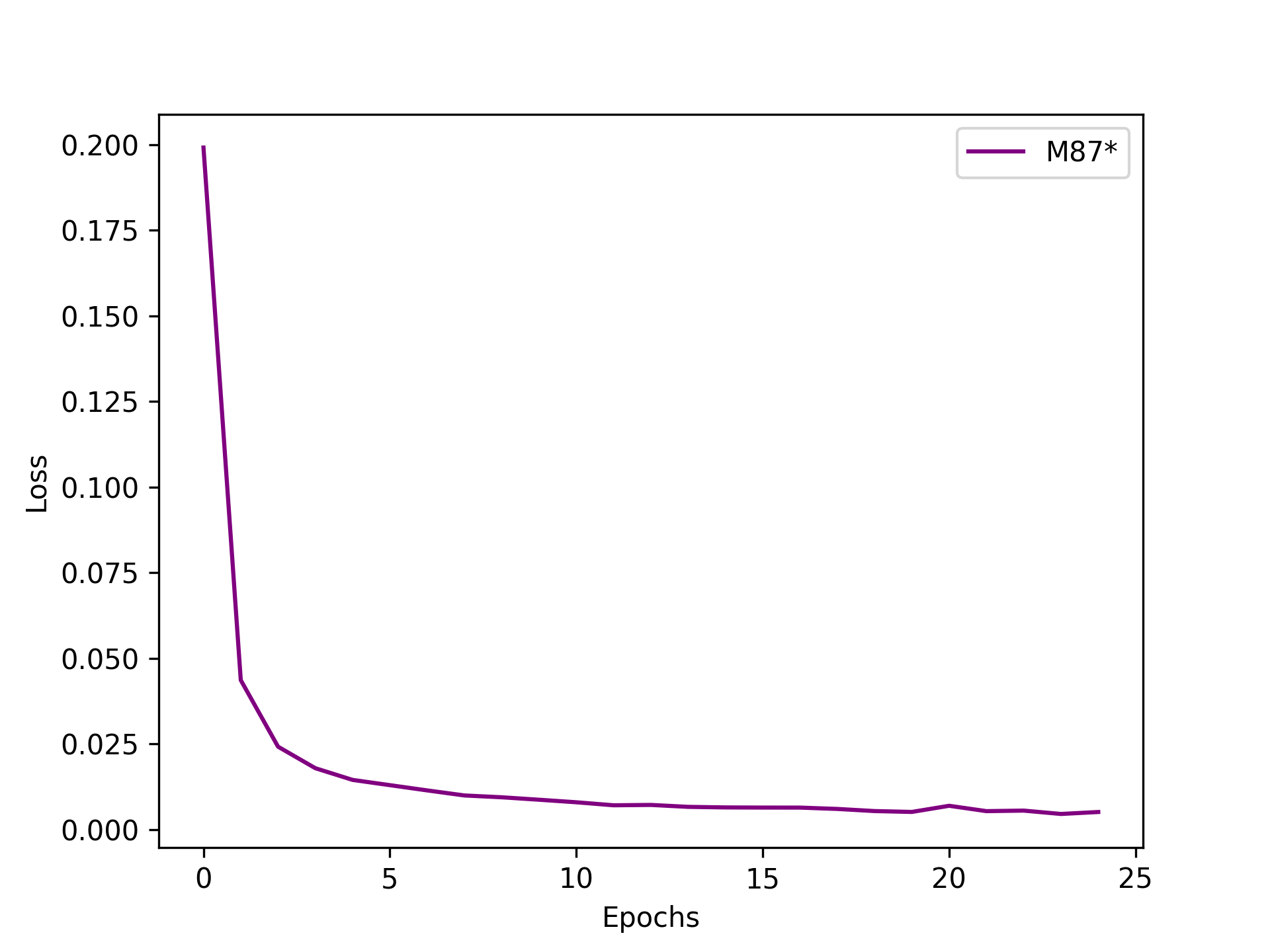}\hspace{1mm}
\includegraphics[scale=0.32]{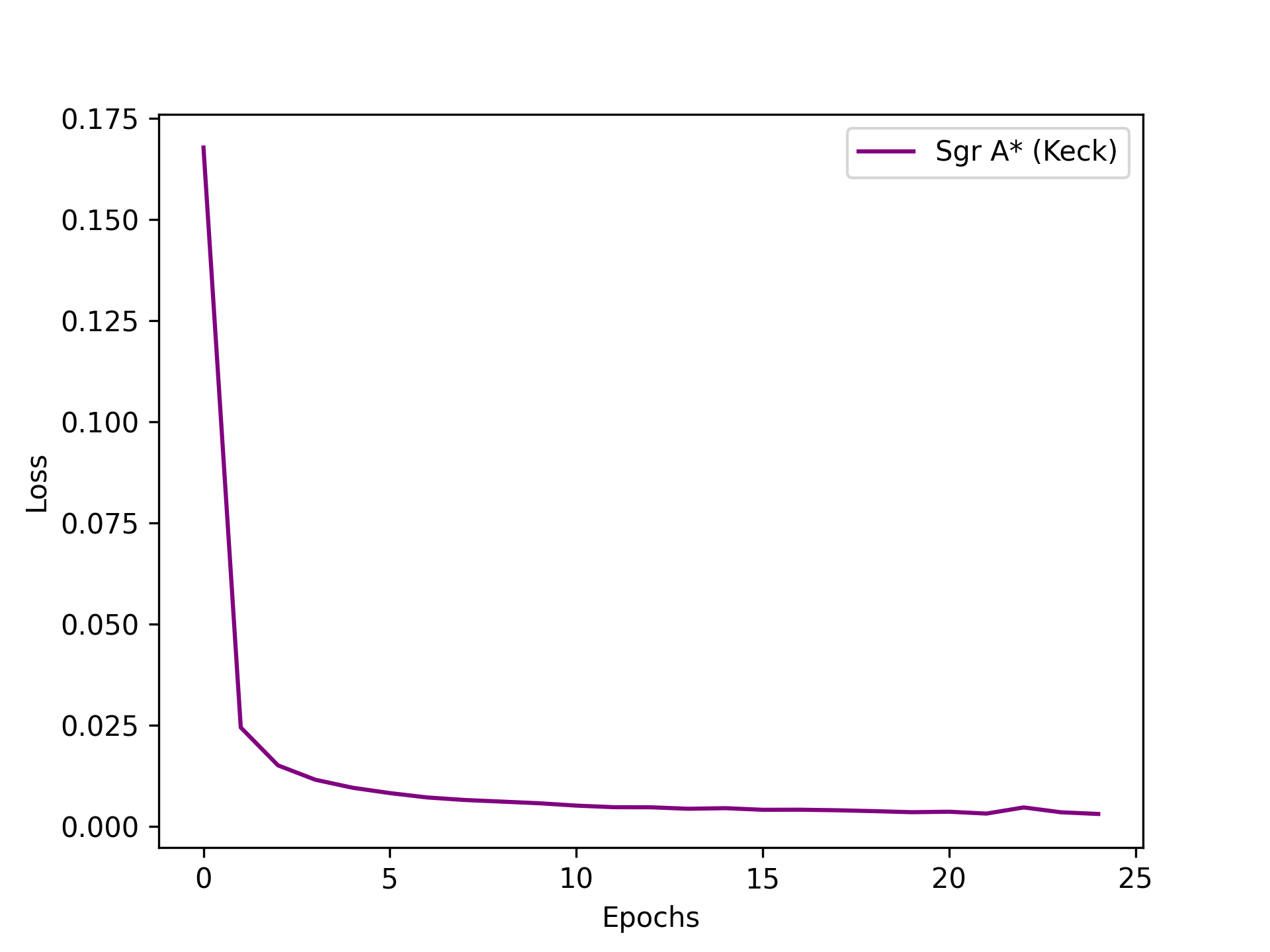}\hspace{1mm}
\includegraphics[scale=0.32]{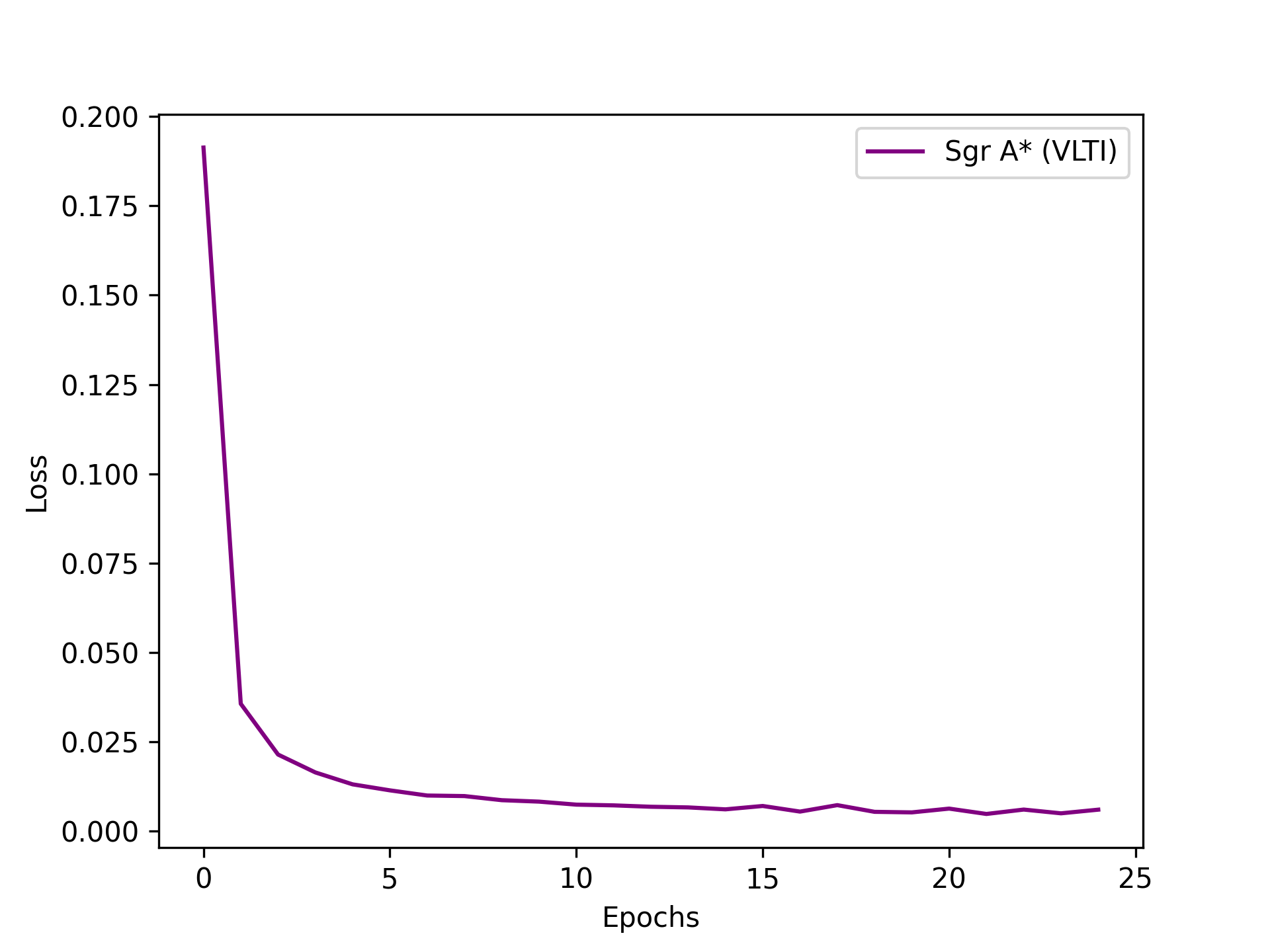}
\caption{Training curves of the FCNN model under the $1-\sigma$ (top row) and $2-\sigma$ (bottom row) observational constraints.}
\label{fig:train1}
\end{figure}

To identify  whether a given parameter configuration provides a black hole shadow compatible with the EHT observational constraints, we employ  FCNN composed of three hidden layers containing 32, 64, and 32 neurons, respectively. Each hidden layer uses the ReLU activation function, while the output layer consists of a single neuron with a sigmoid activation function. This model estimates the probability in which  the input configuration belongs to the class consistent with the EHT observational findings.  The  present network is trained using the Adam optimizer combined with the binary cross-entropy loss function, being  well-suited to the binary classification problem.  The independent test set is exclusively used to evaluate the performance and  the generalization ability of the trained classifier. The corresponding training and  the validation loss curves are shown in Fig.~(\ref{fig:train1}), for the $1-\sigma$ and $2-\sigma$ observation constraints.

As shown in Fig.~(\ref{fig:train1}), the training and the  validation losses decrease rapidly during the initial epochs before gradually reaching stable values. This aspect reveals  efficient learning and fast convergence of the optimization process.  The strong correlation between the training and  the validation loss curves indicates a  good generalization ability, with no significant overfitting. The subtle fluctuations observed in the validation loss are expected due to the stochastic nature of mini-batch optimization and do not affect the model  global convergence. The trained FCNN  has been subsequently  examined   with the help of  an independent test set.  For each observational dataset, the classification accuracy  has been  computed separately.  The obtained results   show  that the classifier reliably distinguishes between the  parameter regions  generating  models  being compatible and incompatible with the EHT observational constraints. This  reveals  that the network has successfully learned the underlying decision boundary in the reduced moduli  space  $(Q,\gamma)$.  For $M87^*$, $SgrA^*_{\mathrm{VLTI}}$, and $SgrA^*_{\mathrm{Keck}}$ black hole   models,   the  corresponding  performance metrics for the $1\sigma$ and $2\sigma$ observational constraints are illustrated  in Table~(\ref{tab1}).

\begin{table}[!ht]

\centering

\begin{tabular}{l|c|c}

\toprule

 & $1-\sigma$ & $2-\sigma$ \\

 \midrule

\textbf{Metric} 

& $M87^*$ \quad $SgrA^*_{\mathrm{VLTI}}$ \quad $SgrA^*_{\mathrm{Keck}}$

& $M87^*$ \quad $SgrA^*_{\mathrm{VLTI}}$ \quad $SgrA^*_{\mathrm{Keck}}$ \\

\midrule

\textbf{Test Accuracy} 

& 0.99725 \quad 0.99900 \quad 0.99850

& 0.99800 \quad 0.99925 \quad 0.99825 \\

\bottomrule

\end{tabular}
\caption{Model performance of the FCNN model for the $1-\sigma$ and $2-\sigma$ observational constraints.}
\label{tab1}
\end{table}

To examine  the robustness of the trained classifier, we use a Monte Carlo voting strategy based on  the Gaussian perturbations of the normalized input features. For each sample in the independent test set, 20 perturbed realizations are generated by considering   a Gaussian noise with a standard deviation of  $0.01$  to the normalized input vector. Each realization is processed independently by the FCNN. The final prediction is obtained  by  using a majority vote among the 20 outputs. Since the perturbations are  implemented  only after training, this procedure does not affect the ground-truth labels produced by exploiting  the parallel programing   computations. However, it  permits  to measure the classifier  stability in the vicinity of each test sample. Furthermore, it reduces the classifier  sensitivity to small numerical perturbations. The corresponding voting accuracies are presented in Table~(\ref{tab1}). Despite  the compatibility labels being derived from  the deterministic comparison between the numerically calculated black hole shadow and the observational limits of the  EHT data, the FCNN should not be intended to replace the underlying physical simulations. However,   it serves as a computational surrogate model approximating  the decision boundary learned from  dataset built from parallel programing computations. Once trained, the network rapidly predicts whether a given point in the parameter space satisfies the EHT observational constraints,  avoiding repeated numerical shadow computations during large-scale moduli space  explorations.  For  $M87^*$, $SgrA^*_{\mathrm{VLTI}}$, and $SgrA^*_{\mathrm{Keck}}$ black  hole   models,  the confusion matrices corresponding to the $1$--$\sigma$ and $2$--$\sigma$ observational constraints are  represented  in Table~(\ref{c11}). 

\begin{table}[!ht]
\centering

\begin{tabular}{l|l|c|c|c}
\toprule
\textbf{} & {Class} & {M87$^*$} 
& {SgrA$^*_{\mathrm{Keck}}$} & {SgrA$^*_{\mathrm{VLTI}}$} \\
\midrule
\multirow{2}{*}{$1-\sigma$} 
&  & Class 0 \quad Class 1 &  Class 0 \quad Class 1 &  Class 0 \quad Class 1 \\
& Class 0 &1600 \qquad  4 &1734 \qquad 3 &2283\qquad 1 \\
& Class 1 &4  \qquad 2392 &0 \qquad 2263 & 6 \qquad1711 \\
\midrule
\multirow{2}{*}{$2-\sigma$} 
&  & Class 0 \quad Class 1 &  Class 0 \quad Class 1 &  Class 0 \quad Class 1 \\
& Class 0 &389 \qquad  10 &799 \qquad 0 &959\qquad 0 \\
& Class 1 &1 \qquad 3600 &4 \qquad 3197 &6 \qquad 3035 \\
\bottomrule
\end{tabular}
\caption{Confusion matrix using the FCNN model  for  $1-\sigma$ and $2-\sigma$ observational constraint.}
\label{c11}
\end{table}

These results reveal that the FCNN correctly classifies the vast majority of samples, with only a very small number of false positives and false negatives.  The consistently high values for precision, recall, F$_1$ score, and AUC confirm the robustness, reliability, and discriminatory power of the proposed alternative model applied to all observational datasets.
 A careful examination shows that the classifier performs significantly better under $1$--$\sigma$ observation constraints than under $2$--$\sigma$ constraints. More specifically, the SgrA$^*_{\mathrm{VLTI}}$ dataset consistently yields the best classification accuracy. This difference can be explained by the fact that the $1$--$\sigma$ confidence interval defines a more restrictive region of compatibility. This results in a clearer separation between the two classes. In contrast, the larger $2$--$\sigma$ confidence interval expands the region of the  acceptable parameter  space regions making  the classification task more difficult. Despite this, the model maintains a high level of accuracy across all datasets. This  shows  its robustness and its ability to consider  different confidence levels of the EHT observations.
\section{Conclusion}
Using   machine learning methods assisted by parallel programing  computations, we have  studied  the optical  behaviors  of AdS black holes  in  dRGT-like massive gravity coupled to  the ModMax nonlinear electrodynamics.  Firstly, we have analyzed  the horizon structure through the metric function of  such modified solutions in both the weak and strong massive gravity regimes.  Due to  the complexity of the optical equations, we have  approached  the photon sphere, the black hole shadow, and the energy emission rate. Concretely, we have shown  that   the weak and  the strong massive gravity regimes exhibit distinct optical behaviors. It has been observed, however, that  the electric charge $Q$  has affected only the size of the black hole shadows.  To  establish  a link with  astrophysical observations, we have developed a   numerical code relaying  on parallel programing  computations. Concretely, we have  constrained  the black hole charge using   the EHT  shadow observations.  Then, we  have used such a  resulting constrained parameter space  to generate labeled datasets for a FCNN.  Such a  trained FCNN has successfully classified the  compatible and the  incompatible black hole configurations with accuracies close to or exceeding $99\%$, with the $1$--$\sigma$ datasets having generally yielded slightly more stable performance than the corresponding $2$--$\sigma$ datasets associated with black hole empirical activities. As a result,  we have  revealed   that the combination of  accelerated parallel programing  simulations and machine learning has provided an efficient framework  to  constraint black hole parameters   using the    empirical   shadow optical  results  provided   by  EHT  observational data.  Concretely,  we have discussed   the  black hole parameter regions which could be   compatible with the EHT observations of M87$^*$, SgrA$^*_{\mathrm{VLTI}}$, and SgrA$^*_{\mathrm{Keck}}$ within the $1-\sigma$ and $2-\sigma$ confidence intervals.   Precisely,  we  have found   that the obtained constraints  fit  with the SgrA$^*_{\mathrm{VLTI}}$   observational findings.

This work  comes up  with certain open questions. In future works, for instance,    more advanced machine learning techniques  and additional evaluation metrics, including ROC analysis and precision-recall metrics could be considered.   This may improve the robustness and the  reliability of the proposed framework, as explored in the  study of  the  black hole optical behaviors.  Extended models with modified gravity theories  explored in cosmological model scenarios where coupling terms could bring new features in black holes  which could be  supported by string theory and D-brane physics could be also developed  \cite{mod1}. We hope to cover these  issues and related topics in future investigations.\\\

{\bf Data availability}\\
  The data are available from the corresponding author upon reasonable request. \\

\section*{Acknowledgements}   
MJ  would like to thank S. E. Baddis and  H. Belmahi  for  recent  collaborations  and discussions on numerical computations.  She  gratefully acknowledges the financial support of the CNRST in the frame 
of the PhD Associate Scholarship Program PASS.

\end{document}